\documentclass[aps,twocolumn,prb,amsmath,amssymb,showpacs,notitlepage,superscriptaddress]{revtex4-1}
\usepackage{graphicx}% Include figure files
\usepackage{dcolumn}% Align table columns on decimal point
\usepackage{bm}% bold math
\usepackage{wrapfig}
\usepackage{amsmath,amsfonts,amssymb,bm}
\usepackage[usenames]{color}
\usepackage{amsmath}
\begin{document}

\title{Antiferromagnetic anisotropy determination by spin Hall magnetoresistance}

\author{Hua Wang}
\affiliation{State Key Laboratory of Surface Physics and Department of Physics, Fudan University, Shanghai 200433, China}
\affiliation{Collaborative Innovation Center of Advanced Microstructures, Fudan University, Shanghai 200433, China}
\affiliation{Institute for Materials Research, Tohoku University, Sendai 980-8577, Japan}
\author{Dazhi Hou}
\email{dazhi.hou@imr.tohoku.ac.jp}\affiliation{WPI Advanced Institute for Materials Research, Tohoku University, Sendai 980-8577, Japan}
\author{Zhiyong Qiu}
\affiliation{WPI Advanced Institute for Materials Research, Tohoku University, Sendai 980-8577, Japan}
\author{Takashi Kikkawa}
\affiliation{Institute for Materials Research, Tohoku University, Sendai 980-8577, Japan}
\affiliation{WPI Advanced Institute for Materials Research, Tohoku University, Sendai 980-8577, Japan}
\author{Eiji Saitoh}
\affiliation{Institute for Materials Research, Tohoku University, Sendai 980-8577, Japan}
\affiliation{WPI Advanced Institute for Materials Research, Tohoku University, Sendai 980-8577, Japan}
\affiliation{Advanced Science Research Center, Japan Atomic Energy Agency, Tokai 319-1195, Japan}
\author{Xiaofeng Jin}
\affiliation{State Key Laboratory of Surface Physics and Department of Physics, Fudan University, Shanghai 200433, China}
\affiliation{Collaborative Innovation Center of Advanced Microstructures, Fudan University, Shanghai 200433, China}

\begin{abstract}
An electric method for measuring magnetic anisotropy in antiferromagnetic insulators (AFI) is proposed. When a metallic film with strong spin-orbit interaction, e.g. platinum (Pt), is deposited on an AFI, its resistance should be affected by the direction of the AFI N\'{e}el vector due to the spin Hall magnetoresistance (SMR). Accordingly, the direction of the AFI N\'{e}el vector, which is affected by both the external magnetic field and the magnetic anisotropy, is reflected in resistance of Pt. The magnetic field angle dependence of the resistance of Pt on AFI is calculated by considering the SMR, which indicates that the antiferromagnetic anisotropy can be obtained experimentally by monitoring the Pt resistance in strong magnetic fields. Calculations are performed for realistic systems such as Pt/Cr$_{2} $O$_{3} $, Pt/NiO and Pt/CoO.
\end{abstract}
\date{\today}
\maketitle

\section{Introduction}

Antiferromagnetic materials, which were passively used to pin the magnetization of the adjacent magnetic layer through exchange bias, have now been gaining renewed attention due to the emerging antiferromagnetic spintronics.\cite{Jungwirth,Seki,Wu,Shang,Hou,Lin,Manchon,wadley,marti,kriegner,hahn,qiu,jhhan,hoogeboom} Recently, the realization of all electric writing and readout antiferromagnetic (AFM)  solid-state memory shows the efficient approach for manipulating AFM moments,\cite{wadley} which is well beyond the previous experimental investigation of anisotropic magnetoresistance (AMR) in AFMs, \cite{marti,kriegner} and indicates the potential broad application prospects in AFM recording media. Since low power consumption is also an important index for the ultrahigh-density integrated circuit,\cite{hahn} one category of AFM materials, antiferromagnetic insulator (AFI), free of the charge current induced Joule heating because of its insulating nature, appear as promising candidates for future spintronics applications.\cite{Seki,Wu,Shang,qiu,Hou,Lin} Toward the practical application of AFI, it is of fundamental importance to obtain the AFI magnetic anisotropy as it defines the orientation of the N\'{e}el vector $\bm\Delta$ = $\textbf{M}_{\rm{A}}/\rm{M_A}-\textbf{M}_{\rm{B}}/\rm{M_B}$. Owing to the difficulty for \emph{ab initio} calculation of the magnetic anisotropy, \cite{Holzle} experimental measurement provides a unique perspective for the investigation of magnetic anisotropy.

In analogy to the anisotropy determination in ferromagnets, the key point of measuring the antiferromagnetic anisotropy is monitoring the N\'{e}el vector direction under different external magnetic field directions. Generally, the standard approach for probing AFM N\'{e}el vector is X-ray magnetic linear dichroism (XMLD) measurement. For the determination of antiferromagnetic anisotropy, several methods are available based on the fitting results including AMR, \cite{fina,hanchun} magnetic torque, \cite{Gafvert} antiferromagnetic resonance(AFMR),\cite{Foner} Mossbauer spectral \cite{Beckman,Rebbouh} and muon spin relaxation ($\mu$SR) \cite{Rebbouh} study. These methods may work well for AFI bulk materials, however, the measurement usually gets challenging for thin film samples which yield weak signals. Is there any convenient method for the N\'{e}el vector and anisotropy determination in both AFI bulk material and thin films?

Lately, a new type of magnetoresistance (MR) in a normal metal (NM)/ferromagnetic insulator (FI) bilayer systems, so-called spin Hall magnetoresistance (SMR), has drawn intense experimental \cite{Nakayama,Shang} and theoretical \cite{Brataas,takahashi,yan} interest. The characteristic of the SMR is that it only depends on the interplay between electron spin polarization $\bm\sigma$ at the NM/FI interface and the magnetization \textbf{M} of FI layer. The SMR, which is defined by the difference of the resistivity for magnetization \textbf{M} perpendicular ($\rho_{\perp}$) and parallel ($\rho_{\parallel}$) to the current $\bm J_{\rm{C}}$, can be formulated as $\rho_{\rm{SMR}}$=$\rho_{\parallel}- \rho_{\perp}$. Since SMR measurement in NM/FI bilayers can directly tell the axis of magnetization \textbf{M} of FI layer without distinguishing the inversion of the magnetization, \cite{Brataas,takahashi,yan} it should be able to determine the antiferromagnetic N\'{e}el vector in a NM/AFI bilayer as well. Besides, in various AFI spintronics experiments, the investigation of spin current transport and SMR when inserting AFI NiO \cite{Shang,Lin,Hou} or CoO \cite{qiu,Lin} between Pt and YIG, indicate the strong interaction between the electron spin polarization $\bm\sigma$ and the AFI N\'{e}el vector $\bm\Delta$. 

In this letter, we calculated SMR in Pt grown on Cr$_{2} $O$_{3} $(110) ,CoO(001) and NiO(001) thin films when rotating the external magnetic field in the film plane. The N\'{e}el vector angle versus external magnetic field direction at different magnetic field magnitudes were investigated systematically. For the uniaxial AFI Cr$_{2}$O$_{3}$, the external field direction dependence of SMR shows different symmetry for magnetic fields below and above the spin-flop field. While for the biaxial AFI NiO and CoO, the external field direction dependence of SMR was only simulated at the magnetic field magnitude higher than the spin-flop field, since even in the single crystal NiO and CoO, there naturally exist two equivalent inplane magnetic domains.\cite{kondoh,uchida1} Meanwhile, we successfully reproduced the anisotropy constant in uniaxial AFI Cr$_{2}$O$_{3}$ by fitting the SMR simulation curve only with the experimental perpendicular susceptibility (which could also be obtained through the first-principle calculation values of exchange interaction constant). This work provides a versatile method to determine the N\'{e}el vector and anisotropy constant for both AFI bulk material and thin films.

\section{SMR in NM/AFI bilayer}

\begin{figure}[t]
  \includegraphics[width=1\linewidth]{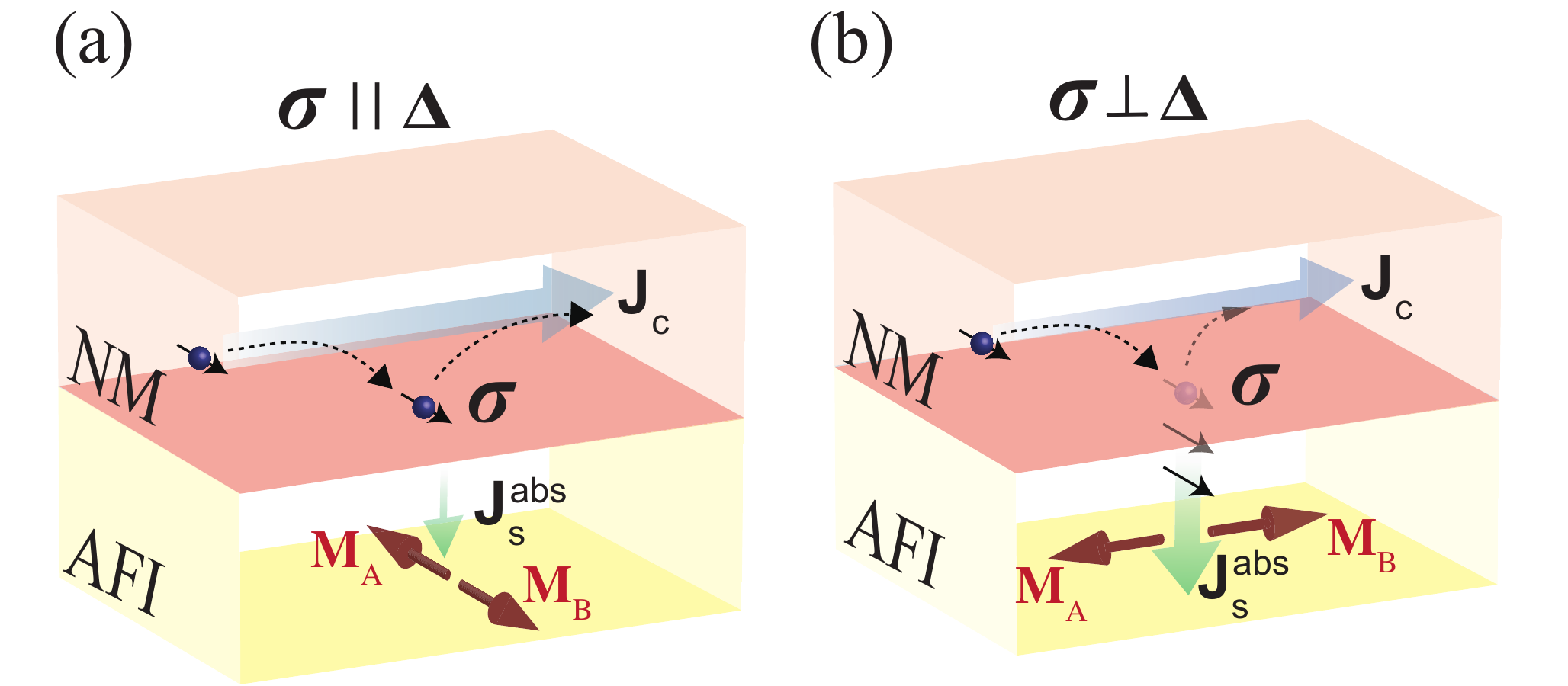}
\caption{(a), (b) Illustrations of the spin Hall magnetoresistance(SMR) in NM (NM=Pt)/AFI (AFI=Cr$_{2}$O$_{3}$, NiO, CoO) bilayer with AFI N\'{e}el vector parallel and perpendicular to the direction of the interface electron spin accumulation. \textbf{J}$_{\rm{C}}$ and \textbf{J}$_{\rm{S}}^{\rm{abs}}$ represent the injected charge current and the spin current absorption in AFI, respectively. \textbf{M}$_{\rm{A}}$ and \textbf{M}$_{\rm{B}}$ are the AFI sublattices.}
\label{fig1}
\end{figure}

Let us consider a NM/AFI bilayer system when an electric current is applied in the Pt film, due to the spin Hall effect (SHE), the charge current will be converted into a spin current  \textbf{J}$_{\rm{S}}$ = $\theta_{\rm{SH}}(\hbar/2e)$\textbf{J}$_{\rm{C}}\times{\bm\sigma}$ with spin polarization  $\bm\sigma$ perpendicular to the electric current \textbf{J}$_{\rm{C}}$.\cite{hirsch,valen,sinova} The spin current with spin polarization  $\bm\sigma$ parallel to the film surface is reflected back and gives rise to a induced charge current due to the inverse spin Hall effect (ISHE),\cite{saitoh,Nakayama} as shown in Figs.~\ref{fig1}(a) and (b). In analogy to the SMR in NM/FI, in NM/AFI bilayers, when electron spin polarization $\bm\sigma$ and N\'{e}el vector $\bm\Delta$ are not parallel, spin-flip scattering is activated. Figure~\ref{fig1}(b) shows when $\bm\sigma$ and $\bm\Delta$ are perpendicular(\textbf{J}$_{\rm{C}}\parallel\bm\Delta$), the spin-transfer torque induced absorption at the NM/AFI interface will be maximized, which gives a higher resistance than the state \textbf{J}$_{\rm{C}}\perp\bm\Delta$. And the conductivity enhancement is expected to be maximized (minimized) when the N\'{e}el vector $\bm\Delta$ is perpendicular (parallel) to \textbf{J}$_{\rm{C}}$. Therefore, the angular dependence measurement of SMR in NM/AFI bilayers can be utilized to determine both the N\'{e}el vector and anisotropy constant in AFI.

In an AFM material, the N\'{e}el vector will stay along the easy axis below the N\'{e}el temperature due to the anisotropy. When applying magnetic field  \textbf{H} parallel to the easy axis with magnitude larger than the critical field H$_{\rm{C}}$, the N\'{e}el vector will suddenly changes its direction perpendicular to \textbf{H}, this first-order transition is called spin-flop transition. Since in general cases, the N\'{e}el vector in AFM is determined by both the external magnetic field and the magnetic anisotropy, a natural question is that if magnetic field  \textbf{H} deviates from the easy axis with angle $\theta_{\rm{H}}$, which direction should the AFM N\'{e}el vector $\bm\Delta$ point to? Kittel,\cite{Kittel} Keffer and Kittel,\cite{Keffer} Nagamiya,\cite{Nagamiya} and others \cite{Nagamiya2} have treated the dynamic response of antiferromagnetically coupled sublattices under different magnetic field direction with molecular field approximation, and here we only focus on the static equilibrium condition.

\begin{figure}[t]
  \includegraphics[width=0.6\linewidth]{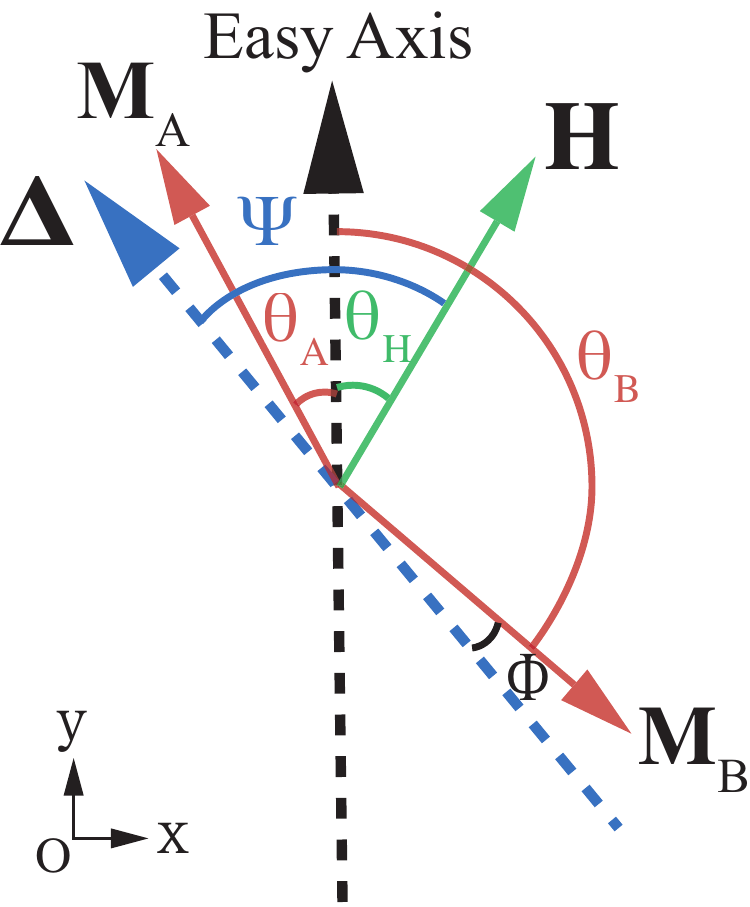}
\caption{Schematic for the distribution of the sublattice magnetizations in uniaxial AFI Cr$_{2}$O$_{3}$(110) thin film and the external magnetic field \textbf{H}. The easy axis and the N\'{e}el vector $\bm\Delta$ in Cr$_{2}$O$_{3}$(110) are represented with the dashed arrows. The angles relative to the easy axis are labeled with $\theta_{\rm{A}}, \theta_{\rm{B}}$ and $\theta_{\rm{H}}$. $\phi$ and $\psi$ represent the tilting angle, the angle between the external field and the N\'{e}el vector, respectively.}
\label{fig2}
\end{figure}

The total magnetic energy of a bulk collinear two-sublattice AFM in a external magnetic field can be phenomenologically written in the following form\cite{Bogdanov}:

\begin{eqnarray}\label{eq1}
E_{\rm{tot}}&=&\int w(\textbf{M}_{\rm{A}},\textbf{M}_{\rm{B}}) dV\nonumber\\ 
&=&\int\{J\textbf{M}_{\rm{A}}\cdot\textbf{M}_{\rm{B}}-\textbf{H}\cdot(\textbf{M}_{\rm{A}}+\textbf{M}_{\rm{B}})+\varepsilon_{\rm{ani}}\} dV 
\end{eqnarray}

\noindent where \textbf{M}$_{\rm{i}}$($\rm{i}$=A,B), $J(J>0)$, \textbf{H} and $\varepsilon_{\rm{ani}}$ represent the sublattice magnetization, exchange interaction constant, external magnetic field and magnetic anisotropy energy respectively. 

\subsection{Uniaxial AFI Cr$_{2}$O$_{3}$}

In uniaxial AFI Cr$_{2}$O$_{3}$(110) film, the vectors \textbf{M}$_{\rm{A}}$, \textbf{M}$_{\rm{B}}$, \textbf{H} in Eq.~\eqref{eq1} and the corresponding angles  $\theta_{\rm{A}}, \theta_{\rm{B}}, \theta_{\rm{H}}$ with the easy axis  are illustrated in Fig.~\ref{fig2}.  $\psi$ is the angle between the external magnetic field and the N\'{e}el vector. The magnetic anisotropy energy could be represented as $\varepsilon_{\rm{ani}}=-\frac{K}{2}(\rm{cos}^2\theta_{\rm{A}}+\rm{cos}^2\theta_{\rm{B}})$, $K (K>0)$ is the anisotropy constant. In Fig.~\ref{fig2} the external magnetic field  \textbf{H} can be decomposed into components parallel and perpendicular to N\'{e}el vector $\bm\Delta$. The parallel component magnetizes  \textbf{M}$_{\rm{A}}$ with changing the magnitude from  M$_{0}$ to M$_{0}$+$\frac{1}{2}\chi_{\parallel}$H$_{\parallel}$, here  $\chi_{\parallel}$ represents parallel susceptibility with \textbf{H} $\parallel\bm\Delta$. While the perpendicular component drives antiferromagnetic ordered spins to tilt a small angle $\phi$ from the N\'{e}el vector direction, for Cr$_{2} $O$_{3} $ the angle $\phi$ is only 1.5$^\circ$ when \textbf{H} $\perp\bm\Delta$ with magnitude H $\approx$ 60 kOe.\cite{Foner} Assuming the  $\chi_{\parallel}^2$ part is negligibly small, we can get the balance of torque equations from Eq.~\eqref{eq1}, written as \cite{Nagamiya2}

\begin{gather}
(\rm{M}_{0}+\frac{1}{2}\chi_{\parallel}\rm{H}_{\parallel}) \rm{H} sin(\psi-\phi)- \it{J} \rm{M}_{0}^2\rm{sin} 2\phi- \it{K} \rm{cos} \theta_{\rm{A}}\rm{sin} \theta_{\rm{A}}=0\nonumber\\
(\rm{M}_{0}-\frac{1}{2}\chi_{\parallel}\rm{H}_{\parallel}) \rm{H} sin(\psi+\phi)- \it{J} \rm{M}_{0}^2\rm{sin} 2\phi+ \it{K} \rm{cos} \theta_{\rm{B}}\rm{sin} \theta_{\rm{B}}=0\nonumber\\
\theta_{\rm{A}}=\theta_{\rm{H}}-\psi+\phi,  \theta_{\rm{B}}=\theta_{\rm{H}}-\psi-\phi+\pi\label{eq2}
\end{gather}

Using the relation 2M$_{0}$sin$\phi$ = $\chi_{\perp}$H$_{\perp}$( $\chi_{\perp}$ is the perpendicular susceptibility with  \textbf{H} $\perp\bm\Delta$) and neglecting the  $\chi_{\parallel}\chi_{\perp}$ part, the results of Eq.~\eqref{eq2} are

\begin{gather}
\chi_{\perp}=\frac{1}{J+(K/2\rm{M}_0^2)cos2(\theta_{\rm{H}}-\psi)}\label{eq3}\\
(\chi_{\perp}-\chi_{\parallel})\rm{H}^2sin\psi cos\psi=\it{K}\rm{sin}2(\psi-\theta_{\rm{H}})\label{eq4}
\end{gather}

Eq.~\eqref{eq3} can be replaced with a simple formula $\chi_{\perp}=\frac{1}{J}$, since the term $(K/2\rm{M}_0^2)cos2(\theta_{\rm{H}}-\psi)$ is usually much smaller than the exchange constant $J$. Therefore, the perpendicular susceptibility $\chi_{\perp}$ is directly related to the exchange constant $J$ (which can be obtained with first-principle calculation \cite{ales,van,pch}) through Eq.~\eqref{eq3}. Define $\theta_{\rm{\Delta}}\equiv\theta_{\rm{H}}-\psi$, and Eq.~\eqref{eq4} can be transformed into

\begin{figure}[t]
  \includegraphics[width=1.15\linewidth]{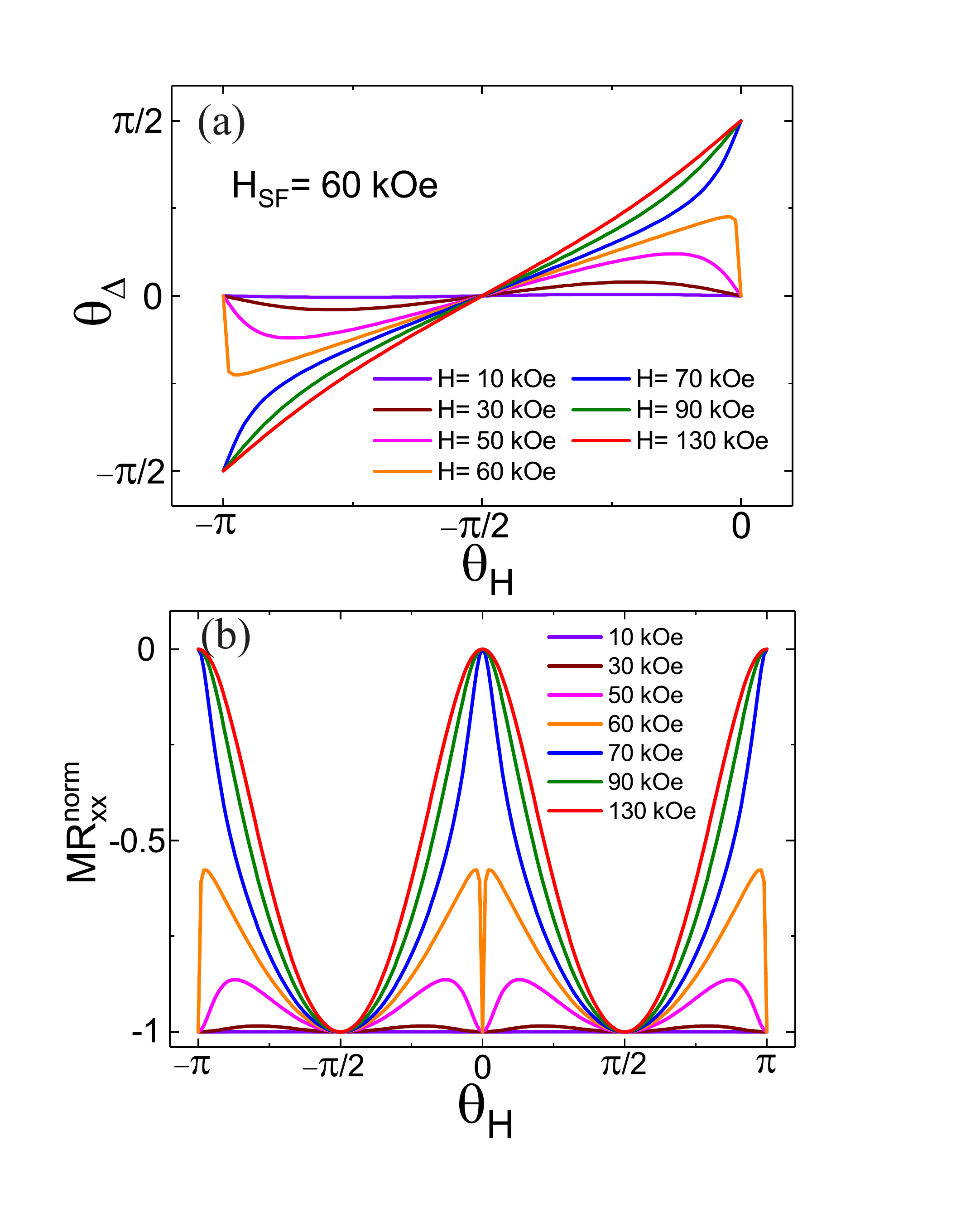}
\vspace{-4em}\caption{(a) Simulation curves of the angular dependence of the N\'{e}el vector in AFI Cr$_{2}$O$_{3}$(110) thin film with different magnetic fields, and the parameters of the bulk material Cr$_{2}$O$_{3}$ at 10 K are utilized for the simulation. (b) The external field dependence of the normalized SMR resistivity in Pt/Cr$_{2}$O$_{3}$(110) are shown with different colors.}
\label{fig3}
\end{figure}

\begin{equation}\label{eq5}
(\chi_{\perp}-\chi_{\parallel})\rm{H}^2sin(\theta_{\rm{H}}-\theta_{\rm{\Delta}})cos(\theta_{\rm{H}}-\theta_{\rm{\Delta}})=-\it{K}\rm{sin}2\theta_{\rm{\Delta}}
\end{equation}

Substituting the parameters in Eq.~\eqref{eq5} with the experimental results of susceptibility $\chi_{\parallel}=1.49\times10^{-6}$ emu/g, $\chi_{\perp}=22.4\times10^{-6}$ emu/g and the anisotropy constant $K$= 38080 ergs/g in bulk material Cr$_{2}$O$_{3}$,\cite{Foner} we plot the $\theta_{\rm{\Delta}}-\theta_{\rm{H}}$ curve under different magnetic field magnitudes, which is shown in Fig.~\ref{fig3}(a). The spin-flop field in AFI Cr$_{2}$O$_{3}$ can be calculated as H$_{\rm{SF}}$ = $\sqrt{2K/(\chi_{\perp}-\chi_{\parallel})}\approx$ 60 kOe at 300 K. In Fig.~\ref{fig3}(a), for H $\ll$ H$_{\rm{SF}}$, the N\'{e}el vector almost stays along the easy axis with just a small perturbation. And this perturbation of sublattice magnetization becomes stronger with increasing the magnitude of the external magnetic field. Especially for the case when H $\approx$ H$_{\rm{SF}}$,  the angle of N\'{e}el vector $\theta_{\rm{\Delta}}$ shows a drastic change when the external magnetic field direction is near the easy axis, since the first order spin-flop transition happens as the magnetic field parallel to the spin axis. While H $\gg$ H$_{\rm{SF}}$, the AFM N\'{e}el vector follows the external field direction with a relative fixied angle $\psi$.

With the external field dependent AFM N\'{e}el vector, it is easier for us to quantitatively analyze the angular dependent SMR in NM/AFI bilayer under fixed magnetic field magnitude. The interface spin current depends on the relative direction of the magnetization and spin accumulation direction, following the formula $(G_{r}/e)$\textbf{m}$\times$(\textbf{m}$\times\bm{\mu}_{\rm{S}}$), where $G_{r}$ is the interface spin-mixing conductance. \textbf{m} is the magnetization direction, and $\bm{\mu}_{\rm{S}}$ is the spin accumulation direction at the interface. \cite{Brataas,Nakayama} In the AFI Cr$_{2}$O$_{3}$, the interface spin accumulation interacts with both the two sublattice magnetizations \textbf{M}$_{\rm{A}}$ and \textbf{M}$_{\rm{B}}$. Since the tilting angle $\phi$ and H$_{\parallel}$ induced magnetization is negligibly small, we can obtain $\bm{\Delta}=\textbf{m}_{\rm{A}}-\textbf{m}_{\rm{B}}\approx2\textbf{m}_{\rm{A}}$. The interface spin current can be described as

\begin{gather}
(G_{r}/e)[\textbf{m}_{\rm{A}}\times(\textbf{m}_{\rm{A}}\times\bm{\mu}_{\rm{S}})+\textbf{m}_{\rm{B}}\times(\textbf{m}_{\rm{B}}\times\bm{\mu}_{\rm{S}})]\nonumber\\ 
=(G_{r}/e)\textbf{m}_{\rm{A}}\times[(\textbf{m}_{\rm{A}}-\textbf{m}_{\rm{B}})\times\bm{\mu}_{\rm{S}}]\nonumber\\ 
=(G_{r}/2e)\bm{\Delta}\times(\bm{\Delta}\times\bm{\mu}_{\rm{S}})\label{eq6}
\end{gather}

\begin{figure}[t]
  \includegraphics[width=0.75\linewidth]{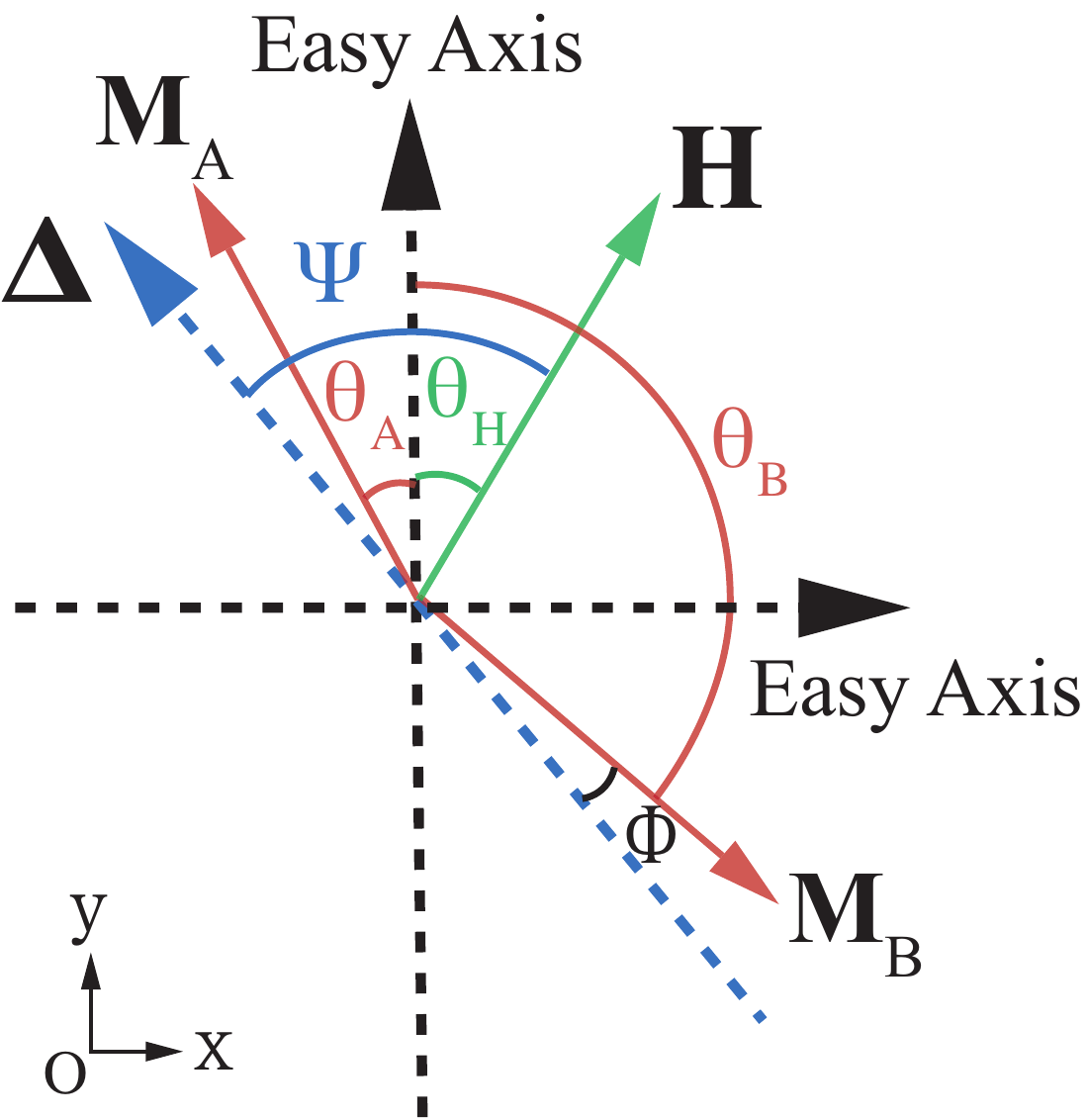}
\caption{Schematic for the distribution of the sublattice magnetizations in biaxial AFI NiO(001), CoO(001) thin film and the external magnetic field \textbf{H}. The two equivalent easy axises and the N\'{e}el vector $\bm\Delta$ in NiO(001), CoO(001) are represented with the dashed arrows. The angles relative to one of the easy axises are labeled with $\theta_{\rm{A}}, \theta_{\rm{B}}$ and $\theta_{\rm{H}}$. $\phi$ and $\psi$ represent the tilting angle, the angle between the external field and the N\'{e}el vector, respectively.}
\label{fig4}
\end{figure}

\noindent where $\textbf{m}_{\rm{i}}=\textbf{M}_{\rm{i}}/\rm{M}_{\rm{i}}$ ($\rm{i}$=A,B) and $\bm\Delta$ represent the sublattice magnetization direction and N\'{e}el vector, respectively. In Eq.~\eqref{eq6}, the AFI N\'{e}el vector $\bm\Delta$ can substitute for the magnetization direction \textbf{m} in the FI. Considering the fact that the current induced spin accumulation is polarized along the easy axis (which we define as the $\bm{y}$ direction), the longitudinal charge current is modulated as  $\bm{y}\cdot[\bm{\Delta}\times(\bm{\Delta}\times\bm{y})]$ and the longitudinal SMR resistivity change is $\rho_{\rm{xx}}=\rho_{\rm{xx0}}- \delta\rho_{\rm{S}}\Delta_{\rm{y}}^2$,  $\delta\rho_{\rm{S}}$ is the term correlating with the spin mixing conductance and $\rho_{\rm{xx0}}$ is the resistivity when magnetic field parallel to the easy axis. In a Pt/YIG bilayer, the ratio between the change of resistivity and the resistivity $\delta\rho_{\rm{S}}/\rho_{\rm{xx0}}$ is about 10$^{-4}$.\cite{Nakayama} For the comparability of angular dependent SMR under different magnetic fields, it is common to use the normalized longitudinal resistivity, which is defined as

\begin{equation}\label{eq7}
\rm{MR}_{\rm{xx}}^{\rm{norm}}=\frac{\rho_{\rm{xx}}-\rho_{\rm{xx0}}}{\delta\rho_{\rm{S}}}=-\Delta_{\rm{y}}^2=-cos^2(\psi-\theta_{\rm{H}})
\end{equation}

Figure~\ref{fig3}(b) shows the simulated angular dependence of normalized longitudinal resistivity in Pt/Cr$_{2}$O$_{3}$(110) under different magnetic fields. $\theta_{\rm{H}}$ is the angle between the easy axis ($\bm{y}$ direction) and the magnetic field, as shown in Fig.~\ref{fig2}. For H $\ll$ H$_{\rm{SF}}$, the curve shows a totally different symmetry with the conventional $\rm{cos}^2\theta$ dependence of SMR measured in NM/AFI. As the external field magnitude increases to the spin-flop field, a notable resistance change appears around $\theta_{\rm{H}}$ = 0, which is also shown in the $\theta_{\rm{\Delta}}-\theta_{\rm{H}}$ curve, due to the competition between the external magnetic field and magnetic anisotropy. With the magnitude increasing far beyond the spin-flop field H $\gg$ H$_{\rm{SF}}$, the SMR curve gradually degenerates into the $\rm{cos}^2\theta$ symmetry.

\subsection{Biaxial AFI NiO and CoO}

\begin{figure}[t]
  \includegraphics[width=1.15\linewidth]{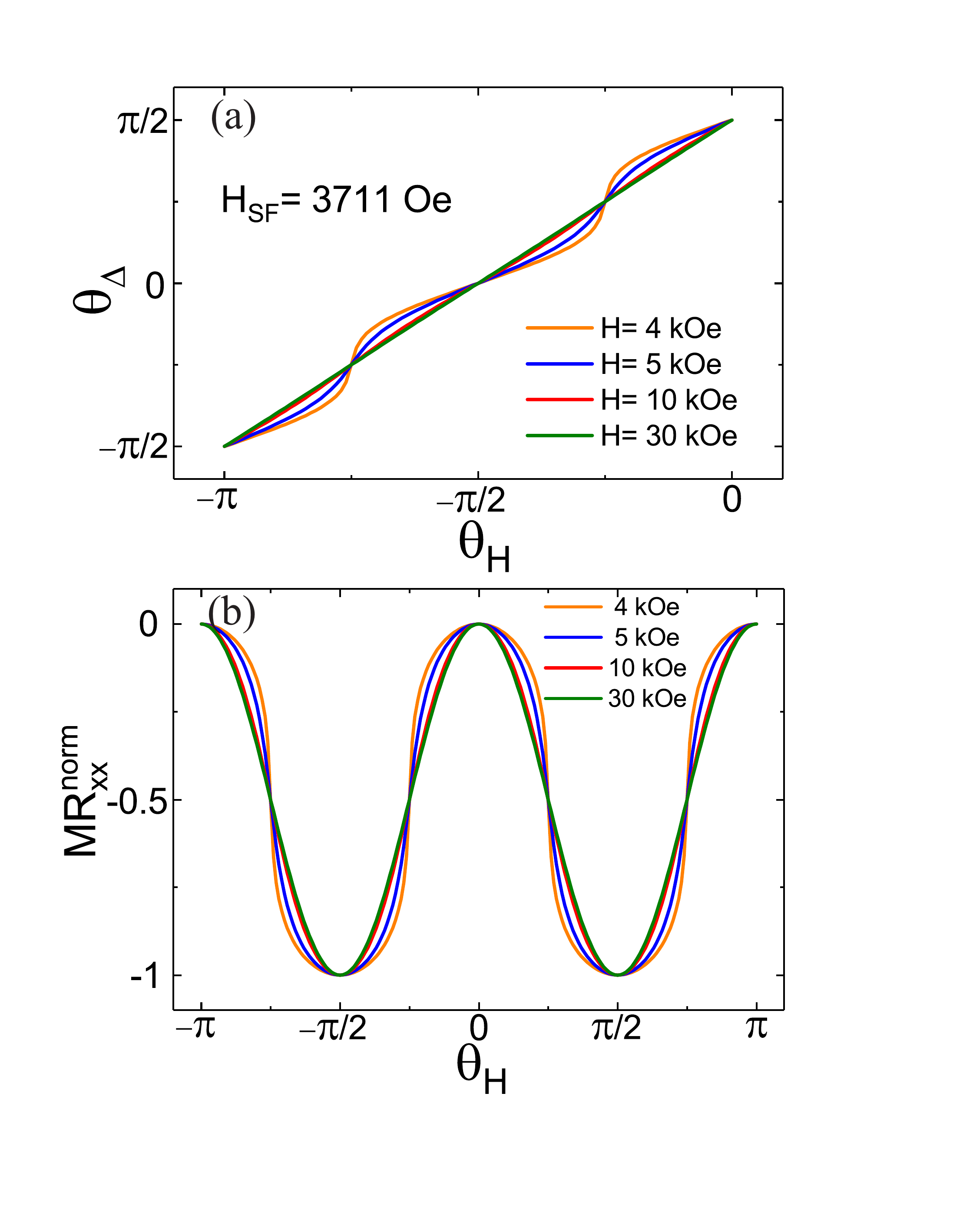}
\vspace{-5em}\caption{(a) Simulation curves of the angular dependence of the N\'{e}el vector in AFI NiO(001) thin film with magnetic field larger than the spin-flop field, and the parameters of the bulk material NiO at 300 K are utilized for the simulation. (b) The corresponding external field dependence of the normalized SMR resistivity in Pt/NiO(001) are shown with different colors.}
\label{fig5}
\end{figure}

For biaxial AFI NiO(001) and CoO(001) films, the magnetic anisotropy energy in Eq.~\eqref{eq1} is represented as $\varepsilon_{\rm{ani}}=-K\rm{cos}4\theta_{\rm{\Delta}}$, and $\theta_{\rm{\Delta}}$ is the angle between N\'{e}el vector $\bm\Delta$ and one of the anisotropy axis. Figure~\ref{fig4} shows the sublattice magnetization vectors and the applied external magnetic field in NiO(001) and CoO(001) plane. The N\'{e}el vector and two equivalent easy axises are represented by the dashed line. We assume the N\'{e}el vector initially stays in the (001) surface plane of NiO and CoO. Following the procedure described above,  we found that $\theta_{\rm{\Delta}}$ satisfies:

\begin{equation}\label{eq8}
(\chi_{\perp}-\chi_{\parallel})\rm{H}^2sin(\theta_{\rm{H}}-\theta_{\rm{\Delta}})cos(\theta_{\rm{H}}-\theta_{\rm{\Delta}})=-4\it{K}\rm{sin}4\theta_{\rm{\Delta}}
\end{equation}

\begin{figure}[t]
  \includegraphics[width=1.15\linewidth]{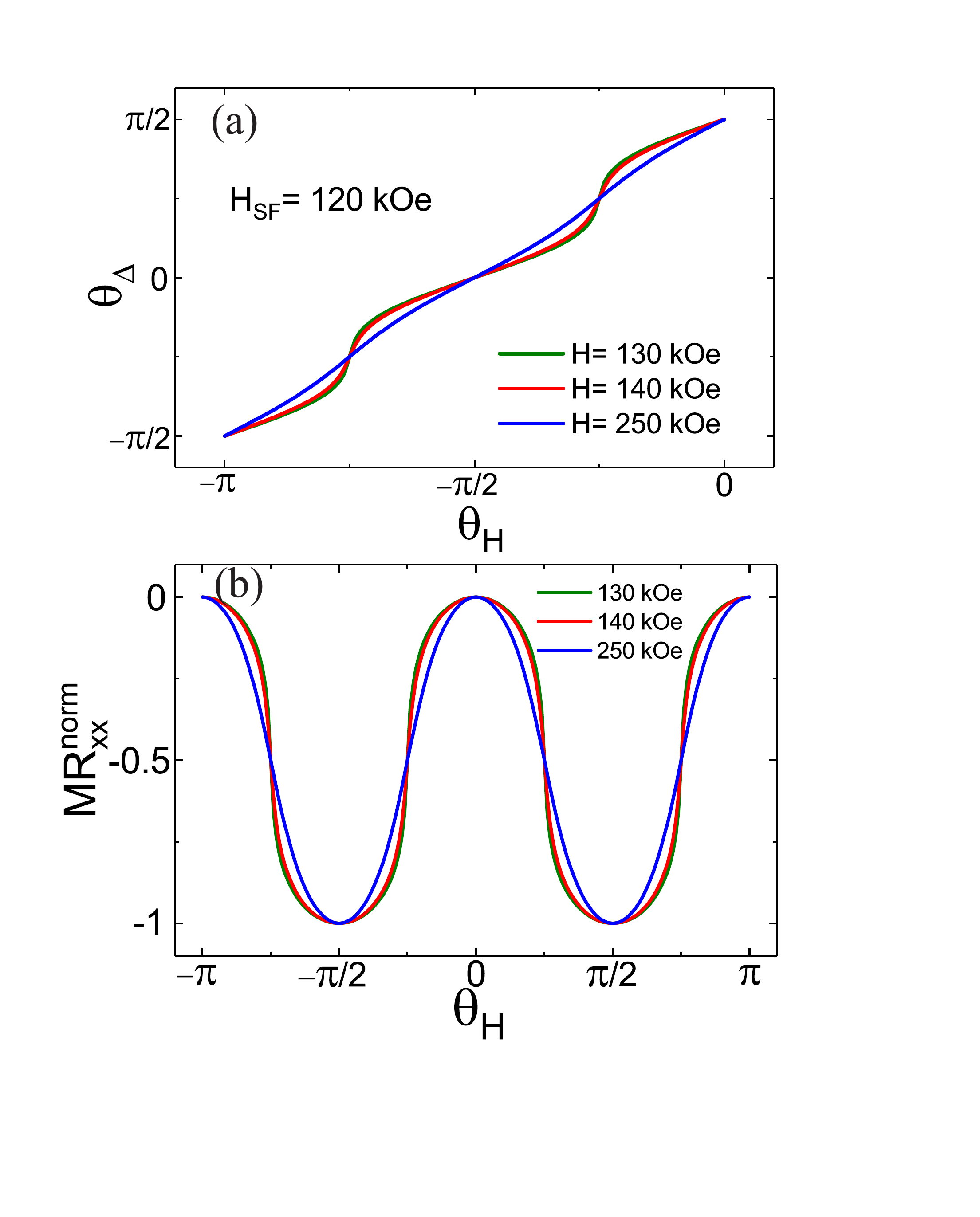}
\vspace{-6em}\caption{(a) Simulation curves of the angular dependence of the N\'{e}el vector in AFI CoO(001) thin film with magnetic field larger than the spin-flop field, and the parameters of the bulk material CoO at 77 K are utilized for the simulation. (b) The corresponding external field dependence of the normalized SMR resistivity in Pt/CoO(001) are shown with different colors.}
\label{fig6}
\end{figure}

In a single-crystalline NiO bulk material, the spin-flop field in the (111) plane is given as H$_{\rm{SF}}$ = $4\sqrt{K/(\chi_{\perp}-\chi_{\parallel})}$ = 2400 Oe at 300 K, while the parallel and perpendicular susceptibility are $\chi_{\parallel}=6.1\times10^{-6}$ emu/g and $\chi_{\perp}=12.2\times10^{-6}$ emu/g, respectively.\cite{uchida2} Combining the magnetic torque data obtained in the (001) and (111) plane of single-crystalline NiO \cite{kondoh} with the  anisotropy constant obtained through the spin-flop field in the (111) plane\cite{uchida2}, we estimate the anisotropy constant in the (001) plane is $K$ = 5.25 ergs/g. Substituting the parameters in Eq.~\eqref{eq8} with the corresponding values in the NiO(001) plane, we obtain the angular dependence of the N\'{e}el vector direction when the external magnetic field rotates in the (001) plane with fixed magnitude, as shown in Fig.~\ref{fig5} (a). When H $\approx$ $H_{\rm{SF}}$,  the distinct bump at the magnetic field direction $\theta_H=\pi/4$ also shows the strong interplay between the external field and anisotropy field. At H $\gg$ H$_{\rm{SF}}$, the N\'{e}el vector $\bm\Delta$ almost follows the magnetic field direction with an angle. The angular dependence of the normalized longitudinal resistivity with different field magnitude is shown in Fig.~\ref{fig5} (b). The angular dependence gradually changes to the $\rm{cos}^2\theta$ type with increasing the magnitude of the external magnetic field.

\begin{figure}[t]
  \includegraphics[width=1\linewidth]{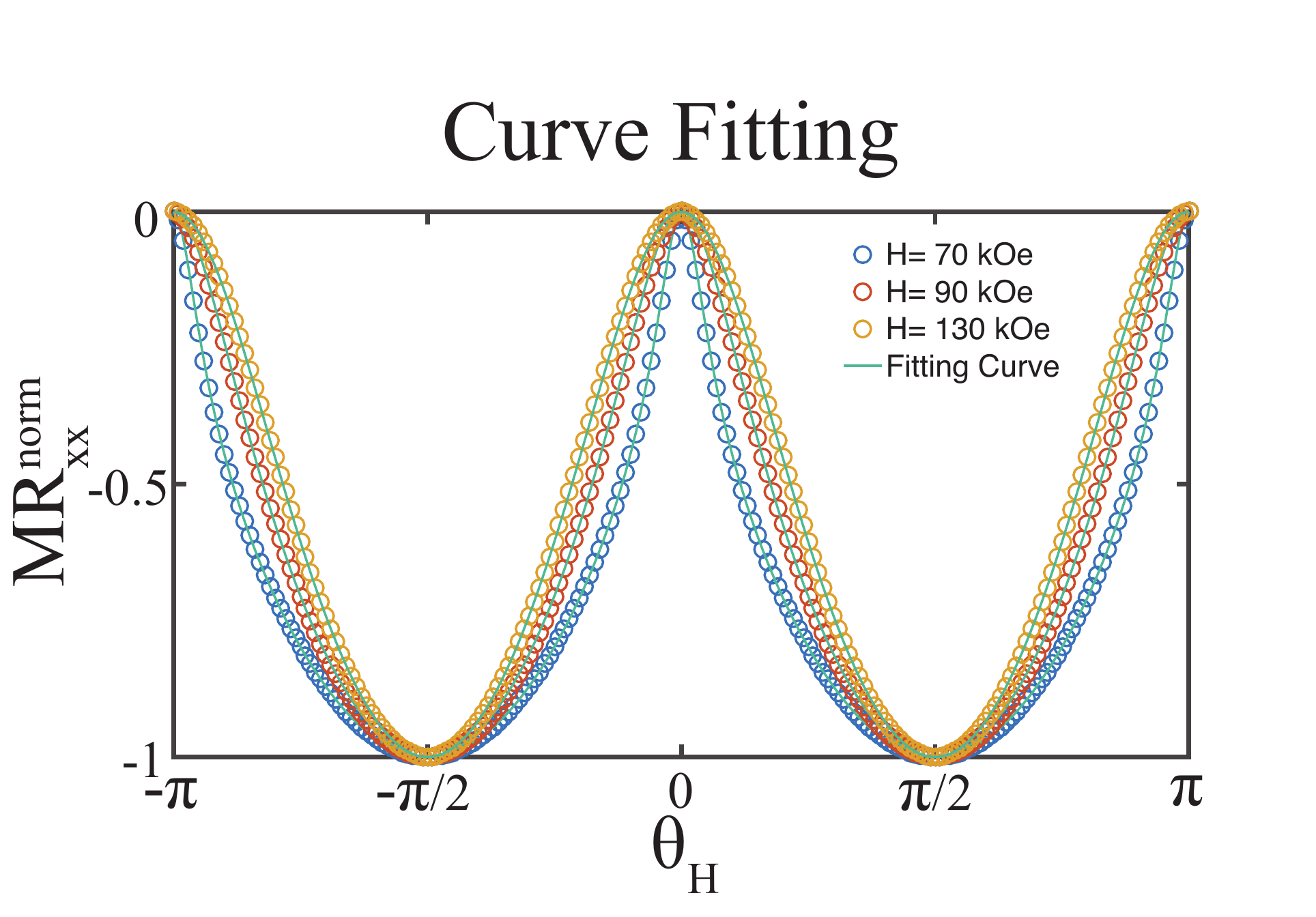}
\caption{The fitting curve of the normalized SMR resistivity in Pt/Cr$_{2}$O$_{3}$(110) at H $>$ H$_{\rm{SF}}$. The open circles represent the simulation results. And the green solid lines are the fitting curves by only taking the perpendicular susceptibility $\chi_{\perp}$.}
\label{fig7}
\end{figure}

In a single-crystalline CoO bulk material, the spin-flop occurs around H$_{\rm{SF}}$ = 120 kOe at 77 K when the external magnetic field is applied along the direction [001]. \cite{inagawa} The corresponding susceptibility values are $\chi_{\parallel}=3.8\times10^{-5}$ emu/g and $\chi_{\perp}=5.35\times10^{-5}$ emu/g respectively. Substituting the parameters in Eq.~\eqref{eq8} with the values in the CoO(001) plane, we simulated the $\theta_{\rm{\Delta}}-\theta_{\rm{H}}$ curve in CoO(001) and MR$_{\rm{xx}}^{\rm{norm}}-\theta_{\rm{H}}$ curve in Pt/CoO(001), which are shown in Figs.~\ref{fig6}(a) and (b). The simulation results are qualitatively consistent with the result obtained in NiO(001).

\section{SMR results fitting}

Since the SMR curve in NM/AFI bilayer shows distinctive line shapes under different magnetic fields, it is possible to obtain the anisotropy constant via the SMR curve fitting. Taking the simulated SMR results of Pt/Cr$_{2}$O$_{3}$(110) as an example, we now try to find out whether the original input parameters could be reproduced through the curve fitting. And the fitting formula is given by combining the Eq.~\eqref{eq5} and Eq.~\eqref{eq7}\cite{Nagamiya2}

\begin{equation}\label{eq9}
\rm{MR}_{\rm{xx}}^{\rm{norm}}=-\rm{cos}^2[\frac{1}{2}\rm{tan}^{-1}(\frac{\rm{sin}2\theta_H}{\rm{cos}2\theta_H-(\chi_{\perp}-\chi_{\parallel})\rm{H}^2/2\it{K}})-\theta_H]
\end{equation}

\noindent where we keep $\chi_{\perp}$ as the known parameter. The $K$ and $\chi_{\parallel}$ are taken as the fitting parameters. Figure~\ref{fig7} shows the original simulated SMR in Pt/Cr$_{2}$O$_{3}$ bilayer with H $>$ H$_{\rm{SF}}$ and a fitting curve. The open circles are the simulation results with the input parameters $\chi_{\parallel}=1.49\times10^{-6}$ emu/g, $\chi_{\perp}=22.4\times10^{-6}$ emu/g, and $K$ = 38080 ergs/g. The green solid lines show the fitting with Eq.~\eqref{eq9} by only taking the experimental value $\chi_{\perp}=22.4\times10^{-6}$ emu/g. The output fitting parameters are $\chi_{\parallel}=1.60\times10^{-6}$ emu/g, and $K$ = 37882 ergs/g, which agrees with the input parameters and proves the feasibility of anisotropy constant determination through SMR measurement in NM/AFI bilayer.

\section{Summary}
In summary, we proposed an electric method for the anisotropy determination in AFIs by using the SMR measurement in NM/AFI bilayer. In both uniaxial and biaxial AFIs, the normalized SMR resistivity in NM/AFI bilayer systems shows different line shapes under different magnetic field magnitudes. Besides, through fitting the results in Pt/Cr$_{2}$O$_{3}$(110),  we obtained the anisotropy constant in Cr$_{2}$O$_{3}$. This new method paves the way for studying both the N\'{e}el vector and anisotropy constant in both AFI bulk material and thin films.

\section*{acknowledgments}

This work is supported by MOST (Grants No. 2015CB921402), NSFC (Grants No. 11374057, No. 11434003 and No. 11421404),  ICC-IMR, Tohoku University, ERATO ``Spin Quantum Rectification Project'' (No. JPMJER1402) from JST, Japan, Grant-in-Aid for Scientific Research on Innovative Area ``Nano Spin Conversion Science'' (No. JP26103005) and Grant-in-Aid for young scientists (B) (No. JP17K14331) from JSPS KAKENHI, Japan. T.K. is supported by JSPS through a research fellowship for young scientists (No. JP15J08026).

\bibliographystyle{apsrev4-1}

\begin{thebibliography}{41}%
\makeatletter
\providecommand \@ifxundefined [1]{%
 \@ifx{#1\undefined}
}%
\providecommand \@ifnum [1]{%
 \ifnum #1\expandafter \@firstoftwo
 \else \expandafter \@secondoftwo
 \fi
}%
\providecommand \@ifx [1]{%
 \ifx #1\expandafter \@firstoftwo
 \else \expandafter \@secondoftwo
 \fi
}%
\providecommand \natexlab [1]{#1}%
\providecommand \enquote  [1]{``#1''}%
\providecommand \bibnamefont  [1]{#1}%
\providecommand \bibfnamefont [1]{#1}%
\providecommand \citenamefont [1]{#1}%
\providecommand \href@noop [0]{\@secondoftwo}%
\providecommand \href [0]{\begingroup \@sanitize@url \@href}%
\providecommand \@href[1]{\@@startlink{#1}\@@href}%
\providecommand \@@href[1]{\endgroup#1\@@endlink}%
\providecommand \@sanitize@url [0]{\catcode `\\12\catcode `\$12\catcode
  `\&12\catcode `\#12\catcode `\^12\catcode `\_12\catcode `\%12\relax}%
\providecommand \@@startlink[1]{}%
\providecommand \@@endlink[0]{}%
\providecommand \url  [0]{\begingroup\@sanitize@url \@url }%
\providecommand \@url [1]{\endgroup\@href {#1}{\urlprefix }}%
\providecommand \urlprefix  [0]{URL }%
\providecommand \Eprint [0]{\href }%
\providecommand \doibase [0]{http://dx.doi.org/}%
\providecommand \selectlanguage [0]{\@gobble}%
\providecommand \bibinfo  [0]{\@secondoftwo}%
\providecommand \bibfield  [0]{\@secondoftwo}%
\providecommand \translation [1]{[#1]}%
\providecommand \BibitemOpen [0]{}%
\providecommand \bibitemStop [0]{}%
\providecommand \bibitemNoStop [0]{.\EOS\space}%
\providecommand \EOS [0]{\spacefactor3000\relax}%
\providecommand \BibitemShut  [1]{\csname bibitem#1\endcsname}%
\let\auto@bib@innerbib\@empty
%</preamble>
\bibitem [{\citenamefont {Jungwirth}\ \emph {et~al.}(2016)\citenamefont
  {Jungwirth}, \citenamefont {Marti}, \citenamefont {Wadley},\ and\
  \citenamefont {Wunderlich}}]{Jungwirth}%
  \BibitemOpen
  \bibfield  {author} {\bibinfo {author} {\bibfnamefont {T.}~\bibnamefont
  {Jungwirth}}, \bibinfo {author} {\bibfnamefont {X.}~\bibnamefont {Marti}},
  \bibinfo {author} {\bibfnamefont {P.}~\bibnamefont {Wadley}}, \ and\ \bibinfo
  {author} {\bibfnamefont {J.}~\bibnamefont {Wunderlich}},\ }\href@noop {}
  {\bibfield  {journal} {\bibinfo  {journal} {Nat. Nanotech.}\ }\textbf
  {\bibinfo {volume} {11}},\ \bibinfo {pages} {231} (\bibinfo {year}
  {2016})}\BibitemShut {NoStop}%
\bibitem [{\citenamefont {Seki}\ \emph {et~al.}(2015)\citenamefont {Seki},
  \citenamefont {Ideue}, \citenamefont {Kubota}, \citenamefont {Kozuka},
  \citenamefont {Takagi}, \citenamefont {Nakamura}, \citenamefont {Kaneko},
  \citenamefont {Kawasaki},\ and\ \citenamefont {Tokura}}]{Seki}%
  \BibitemOpen
  \bibfield  {author} {\bibinfo {author} {\bibfnamefont {S.}~\bibnamefont
  {Seki}}, \bibinfo {author} {\bibfnamefont {T.}~\bibnamefont {Ideue}},
  \bibinfo {author} {\bibfnamefont {M.}~\bibnamefont {Kubota}}, \bibinfo
  {author} {\bibfnamefont {Y.}~\bibnamefont {Kozuka}}, \bibinfo {author}
  {\bibfnamefont {R.}~\bibnamefont {Takagi}}, \bibinfo {author} {\bibfnamefont
  {M.}~\bibnamefont {Nakamura}}, \bibinfo {author} {\bibfnamefont
  {Y.}~\bibnamefont {Kaneko}}, \bibinfo {author} {\bibfnamefont
  {M.}~\bibnamefont {Kawasaki}}, \ and\ \bibinfo {author} {\bibfnamefont
  {Y.}~\bibnamefont {Tokura}},\ }\href@noop {} {\bibfield  {journal} {\bibinfo
  {journal} {Phys. Rev. Lett.}\ }\textbf {\bibinfo {volume} {115}},\ \bibinfo
  {pages} {266601} (\bibinfo {year} {2015})}\BibitemShut {NoStop}%
\bibitem [{\citenamefont {Wu}\ \emph {et~al.}(2016{\natexlab{a}})\citenamefont
  {Wu}, \citenamefont {Zhang}, \citenamefont {KC}, \citenamefont {Borisov},
  \citenamefont {Pearson}, \citenamefont {Jiang}, \citenamefont {Lederman},
  \citenamefont {Hoffmann},\ and\ \citenamefont {BhaWuttacharya}}]{Wu}%
  \BibitemOpen
  \bibfield  {author} {\bibinfo {author} {\bibfnamefont {S.~M.}\ \bibnamefont
  {Wu}}, \bibinfo {author} {\bibfnamefont {W.}~\bibnamefont {Zhang}}, \bibinfo
  {author} {\bibfnamefont {A.}~\bibnamefont {KC}}, \bibinfo {author}
  {\bibfnamefont {P.}~\bibnamefont {Borisov}}, \bibinfo {author} {\bibfnamefont
  {J.~E.}\ \bibnamefont {Pearson}}, \bibinfo {author} {\bibfnamefont {J.~S.}\
  \bibnamefont {Jiang}}, \bibinfo {author} {\bibfnamefont {D.}~\bibnamefont
  {Lederman}}, \bibinfo {author} {\bibfnamefont {A.}~\bibnamefont {Hoffmann}},
  \ and\ \bibinfo {author} {\bibfnamefont {A.}~\bibnamefont {BhaWuttacharya}},\
  }\href@noop {} {\bibfield  {journal} {\bibinfo  {journal} {Phys. Rev. Lett.}\
  }\textbf {\bibinfo {volume} {116}},\ \bibinfo {pages} {097204} (\bibinfo
  {year} {2016}{\natexlab{a}})}\BibitemShut {NoStop}%
\bibitem [{\citenamefont {Shang}\ \emph {et~al.}(2016)\citenamefont {Shang},
  \citenamefont {Zhan}, \citenamefont {Yang}, \citenamefont {Zuo},
  \citenamefont {Xie}, \citenamefont {Liu}, \citenamefont {Zhang},
  \citenamefont {Zhang}, \citenamefont {Li}, \citenamefont {Wang},
  \citenamefont {Wu}, \citenamefont {Zhang},\ and\ \citenamefont {Li}}]{Shang}%
  \BibitemOpen
  \bibfield  {author} {\bibinfo {author} {\bibfnamefont {T.}~\bibnamefont
  {Shang}}, \bibinfo {author} {\bibfnamefont {Q.~F.}\ \bibnamefont {Zhan}},
  \bibinfo {author} {\bibfnamefont {H.~L.}\ \bibnamefont {Yang}}, \bibinfo
  {author} {\bibfnamefont {Z.~H.}\ \bibnamefont {Zuo}}, \bibinfo {author}
  {\bibfnamefont {Y.~L.}\ \bibnamefont {Xie}}, \bibinfo {author} {\bibfnamefont
  {L.~P.}\ \bibnamefont {Liu}}, \bibinfo {author} {\bibfnamefont {S.~L.}\
  \bibnamefont {Zhang}}, \bibinfo {author} {\bibfnamefont {Y.}~\bibnamefont
  {Zhang}}, \bibinfo {author} {\bibfnamefont {H.~H.}\ \bibnamefont {Li}},
  \bibinfo {author} {\bibfnamefont {B.~M.}\ \bibnamefont {Wang}}, \bibinfo
  {author} {\bibfnamefont {Y.~H.}\ \bibnamefont {Wu}}, \bibinfo {author}
  {\bibfnamefont {S.}~\bibnamefont {Zhang}}, \ and\ \bibinfo {author}
  {\bibfnamefont {R.~W.}\ \bibnamefont {Li}},\ }\href@noop {} {\bibfield
  {journal} {\bibinfo  {journal} {Appl. Phys. Lett.}\ }\textbf {\bibinfo
  {volume} {109}},\ \bibinfo {pages} {032410} (\bibinfo {year}
  {2016})}\BibitemShut {NoStop}%
\bibitem [{\citenamefont {Hou}\ \emph {et~al.}(2017)\citenamefont {Hou},
  \citenamefont {Qiu}, \citenamefont {Barker}, \citenamefont {Sato},
  \citenamefont {Yamamoto}, \citenamefont {V\'{e}lez}, \citenamefont
  {Gomez-Perez}, \citenamefont {Hueso}, \citenamefont {Casanova},\ and\
  \citenamefont {Saitoh}}]{Hou}%
  \BibitemOpen
  \bibfield  {author} {\bibinfo {author} {\bibfnamefont {D.}~\bibnamefont
  {Hou}}, \bibinfo {author} {\bibfnamefont {Z.}~\bibnamefont {Qiu}}, \bibinfo
  {author} {\bibfnamefont {J.}~\bibnamefont {Barker}}, \bibinfo {author}
  {\bibfnamefont {K.}~\bibnamefont {Sato}}, \bibinfo {author} {\bibfnamefont
  {K.}~\bibnamefont {Yamamoto}}, \bibinfo {author} {\bibfnamefont
  {S.}~\bibnamefont {V\'{e}lez}}, \bibinfo {author} {\bibfnamefont {J.~M.}\
  \bibnamefont {Gomez-Perez}}, \bibinfo {author} {\bibfnamefont {L.~E.}\
  \bibnamefont {Hueso}}, \bibinfo {author} {\bibfnamefont {F.}~\bibnamefont
  {Casanova}}, \ and\ \bibinfo {author} {\bibfnamefont {E.}~\bibnamefont
  {Saitoh}},\ }\href@noop {} {\bibfield  {journal} {\bibinfo  {journal} {Phys.
  Rev. Lett.}\ }\textbf {\bibinfo {volume} {118}},\ \bibinfo {pages} {147202}
  (\bibinfo {year} {2017})}\BibitemShut {NoStop}%
\bibitem [{\citenamefont {Lin}\ \emph {et~al.}(2016)\citenamefont {Lin},
  \citenamefont {Chen}, \citenamefont {Zhang},\ and\ \citenamefont
  {Chien}}]{Lin}%
  \BibitemOpen
  \bibfield  {author} {\bibinfo {author} {\bibfnamefont {W.}~\bibnamefont
  {Lin}}, \bibinfo {author} {\bibfnamefont {K.}~\bibnamefont {Chen}}, \bibinfo
  {author} {\bibfnamefont {S.}~\bibnamefont {Zhang}}, \ and\ \bibinfo {author}
  {\bibfnamefont {C.~L.}\ \bibnamefont {Chien}},\ }\href@noop {} {\bibfield
  {journal} {\bibinfo  {journal} {Phys. Rev. Lett.}\ }\textbf {\bibinfo
  {volume} {116}},\ \bibinfo {pages} {186601} (\bibinfo {year}
  {2016})}\BibitemShut {NoStop}%
\bibitem [{\citenamefont {Manchon}(2017)}]{Manchon}%
  \BibitemOpen
  \bibfield  {author} {\bibinfo {author} {\bibfnamefont {A.}~\bibnamefont
  {Manchon}},\ }\href@noop {} {\bibfield  {journal} {\bibinfo  {journal}
  {physica status solidi (RRL)}\ }\textbf {\bibinfo {volume} {11}},\ \bibinfo
  {pages} {1600409} (\bibinfo {year} {2017})}\BibitemShut {NoStop}%
\bibitem [{\citenamefont {Wadley}\ \emph {et~al.}(2016)\citenamefont {Wadley},
  \citenamefont {Howells}, \citenamefont {\v{Z}elezn\'{y}}, \citenamefont
  {Andrews}, \citenamefont {Hills}, \citenamefont {Campion}, \citenamefont
  {Nov\'{a}k}, \citenamefont {Olejn\'{i}k}, \citenamefont {Maccherozzi},
  \citenamefont {Dhesi}, \citenamefont {Martin}, \citenamefont {Wagner},
  \citenamefont {Wunderlich}, \citenamefont {Freimuth}, \citenamefont
  {Mokrousov}, \citenamefont {Kune\v{s}}, \citenamefont {Chauhan},
  \citenamefont {Grzybowski}, \citenamefont {Rushforth}, \citenamefont
  {Edmonds}, \citenamefont {Gallagher},\ and\ \citenamefont
  {Jungwirth}}]{wadley}%
  \BibitemOpen
  \bibfield  {author} {\bibinfo {author} {\bibfnamefont {P.}~\bibnamefont
  {Wadley}}, \bibinfo {author} {\bibfnamefont {B.}~\bibnamefont {Howells}},
  \bibinfo {author} {\bibfnamefont {J.}~\bibnamefont {\v{Z}elezn\'{y}}},
  \bibinfo {author} {\bibfnamefont {C.}~\bibnamefont {Andrews}}, \bibinfo
  {author} {\bibfnamefont {V.}~\bibnamefont {Hills}}, \bibinfo {author}
  {\bibfnamefont {R.~P.}\ \bibnamefont {Campion}}, \bibinfo {author}
  {\bibfnamefont {V.}~\bibnamefont {Nov\'{a}k}}, \bibinfo {author}
  {\bibfnamefont {K.}~\bibnamefont {Olejn\'{i}k}}, \bibinfo {author}
  {\bibfnamefont {F.}~\bibnamefont {Maccherozzi}}, \bibinfo {author}
  {\bibfnamefont {S.~S.}\ \bibnamefont {Dhesi}}, \bibinfo {author}
  {\bibfnamefont {S.~Y.}\ \bibnamefont {Martin}}, \bibinfo {author}
  {\bibfnamefont {T.}~\bibnamefont {Wagner}}, \bibinfo {author} {\bibfnamefont
  {J.}~\bibnamefont {Wunderlich}}, \bibinfo {author} {\bibfnamefont
  {F.}~\bibnamefont {Freimuth}}, \bibinfo {author} {\bibfnamefont
  {Y.}~\bibnamefont {Mokrousov}}, \bibinfo {author} {\bibfnamefont
  {J.}~\bibnamefont {Kune\v{s}}}, \bibinfo {author} {\bibfnamefont {J.~S.}\
  \bibnamefont {Chauhan}}, \bibinfo {author} {\bibfnamefont {M.~J.}\
  \bibnamefont {Grzybowski}}, \bibinfo {author} {\bibfnamefont {A.~W.}\
  \bibnamefont {Rushforth}}, \bibinfo {author} {\bibfnamefont {K.~W.}\
  \bibnamefont {Edmonds}}, \bibinfo {author} {\bibfnamefont {B.~L.}\
  \bibnamefont {Gallagher}}, \ and\ \bibinfo {author} {\bibfnamefont
  {T.}~\bibnamefont {Jungwirth}},\ }\href@noop {} {\bibfield  {journal}
  {\bibinfo  {journal} {Science}\ }\textbf {\bibinfo {volume} {351}},\ \bibinfo
  {pages} {587} (\bibinfo {year} {2016})}\BibitemShut {NoStop}%
\bibitem [{\citenamefont {Marti}\ \emph {et~al.}(2014)\citenamefont {Marti},
  \citenamefont {Fina}, \citenamefont {Frontera}, \citenamefont {Liu},
  \citenamefont {Wadley}, \citenamefont {He}, \citenamefont {Paull},
  \citenamefont {Clarkson}, \citenamefont {Kudrnovsk\'{y}}, \citenamefont
  {Turek}, \citenamefont {Kune\v{s}}, \citenamefont {Yi}, \citenamefont {Chu},
  \citenamefont {Nelson}, \citenamefont {You}, \citenamefont {Arenholz},
  \citenamefont {Salahuddin}, \citenamefont {Fontcuberta}, \citenamefont
  {Jungwirth},\ and\ \citenamefont {Ramesh}}]{marti}%
  \BibitemOpen
  \bibfield  {author} {\bibinfo {author} {\bibfnamefont {X.}~\bibnamefont
  {Marti}}, \bibinfo {author} {\bibfnamefont {I.}~\bibnamefont {Fina}},
  \bibinfo {author} {\bibfnamefont {C.}~\bibnamefont {Frontera}}, \bibinfo
  {author} {\bibfnamefont {J.}~\bibnamefont {Liu}}, \bibinfo {author}
  {\bibfnamefont {P.}~\bibnamefont {Wadley}}, \bibinfo {author} {\bibfnamefont
  {Q.}~\bibnamefont {He}}, \bibinfo {author} {\bibfnamefont {R.~J.}\
  \bibnamefont {Paull}}, \bibinfo {author} {\bibfnamefont {J.~D.}\ \bibnamefont
  {Clarkson}}, \bibinfo {author} {\bibfnamefont {J.}~\bibnamefont
  {Kudrnovsk\'{y}}}, \bibinfo {author} {\bibfnamefont {I.}~\bibnamefont
  {Turek}}, \bibinfo {author} {\bibfnamefont {J.}~\bibnamefont {Kune\v{s}}},
  \bibinfo {author} {\bibfnamefont {D.}~\bibnamefont {Yi}}, \bibinfo {author}
  {\bibfnamefont {J.-H.}\ \bibnamefont {Chu}}, \bibinfo {author} {\bibfnamefont
  {C.~T.}\ \bibnamefont {Nelson}}, \bibinfo {author} {\bibfnamefont
  {L.}~\bibnamefont {You}}, \bibinfo {author} {\bibfnamefont {E.}~\bibnamefont
  {Arenholz}}, \bibinfo {author} {\bibfnamefont {S.}~\bibnamefont
  {Salahuddin}}, \bibinfo {author} {\bibfnamefont {J.}~\bibnamefont
  {Fontcuberta}}, \bibinfo {author} {\bibfnamefont {T.}~\bibnamefont
  {Jungwirth}}, \ and\ \bibinfo {author} {\bibfnamefont {R.}~\bibnamefont
  {Ramesh}},\ }\href@noop {} {\bibfield  {journal} {\bibinfo  {journal} {Nat.
  Mater.}\ }\textbf {\bibinfo {volume} {13}},\ \bibinfo {pages} {367} (\bibinfo
  {year} {2014})}\BibitemShut {NoStop}%
\bibitem [{\citenamefont {Kriegner}\ \emph {et~al.}(2016)\citenamefont
  {Kriegner}, \citenamefont {V\'{y}born\'{y}}, \citenamefont {Olejn\'{i}k},
  \citenamefont {Reichlov\'{a}}, \citenamefont {Nov\'{a}k}, \citenamefont
  {Marti}, \citenamefont {Gazquez}, \citenamefont {Saidl}, \citenamefont
  {N\v{e}mec}, \citenamefont {Volobuev}, \citenamefont {Springholz},
  \citenamefont {Hol\'{y}},\ and\ \citenamefont {Jungwirth}}]{kriegner}%
  \BibitemOpen
  \bibfield  {author} {\bibinfo {author} {\bibfnamefont {D.}~\bibnamefont
  {Kriegner}}, \bibinfo {author} {\bibfnamefont {K.}~\bibnamefont
  {V\'{y}born\'{y}}}, \bibinfo {author} {\bibfnamefont {K.}~\bibnamefont
  {Olejn\'{i}k}}, \bibinfo {author} {\bibfnamefont {H.}~\bibnamefont
  {Reichlov\'{a}}}, \bibinfo {author} {\bibfnamefont {V.}~\bibnamefont
  {Nov\'{a}k}}, \bibinfo {author} {\bibfnamefont {X.}~\bibnamefont {Marti}},
  \bibinfo {author} {\bibfnamefont {J.}~\bibnamefont {Gazquez}}, \bibinfo
  {author} {\bibfnamefont {V.}~\bibnamefont {Saidl}}, \bibinfo {author}
  {\bibfnamefont {P.}~\bibnamefont {N\v{e}mec}}, \bibinfo {author}
  {\bibfnamefont {V.~V.}\ \bibnamefont {Volobuev}}, \bibinfo {author}
  {\bibfnamefont {G.}~\bibnamefont {Springholz}}, \bibinfo {author}
  {\bibfnamefont {V.}~\bibnamefont {Hol\'{y}}}, \ and\ \bibinfo {author}
  {\bibfnamefont {T.}~\bibnamefont {Jungwirth}},\ }\href@noop {} {\bibfield
  {journal} {\bibinfo  {journal} {Nat. Commun.}\ }\textbf {\bibinfo {volume}
  {7}},\ \bibinfo {pages} {11623} (\bibinfo {year} {2016})}\BibitemShut
  {NoStop}%
\bibitem [{\citenamefont {Hahn}\ \emph {et~al.}(2014)\citenamefont {Hahn},
  \citenamefont {de~Loubens}, \citenamefont {Naletov}, \citenamefont {Youssef},
  \citenamefont {Klein},\ and\ \citenamefont {Viret}}]{hahn}%
  \BibitemOpen
  \bibfield  {author} {\bibinfo {author} {\bibfnamefont {C.}~\bibnamefont
  {Hahn}}, \bibinfo {author} {\bibfnamefont {G.}~\bibnamefont {de~Loubens}},
  \bibinfo {author} {\bibfnamefont {V.~V.}\ \bibnamefont {Naletov}}, \bibinfo
  {author} {\bibfnamefont {J.~B.}\ \bibnamefont {Youssef}}, \bibinfo {author}
  {\bibfnamefont {O.}~\bibnamefont {Klein}}, \ and\ \bibinfo {author}
  {\bibfnamefont {M.}~\bibnamefont {Viret}},\ }\href@noop {} {\bibfield
  {journal} {\bibinfo  {journal} {Europhys. Lett.}\ }\textbf {\bibinfo {volume}
  {108}},\ \bibinfo {pages} {57005} (\bibinfo {year} {2014})}\BibitemShut
  {NoStop}%
\bibitem [{\citenamefont {Qiu}\ \emph {et~al.}(2016)\citenamefont {Qiu},
  \citenamefont {Li}, \citenamefont {Hou}, \citenamefont {Arenholz},
  \citenamefont {N'Diaye}, \citenamefont {Tan}, \citenamefont {ichi Uchida},
  \citenamefont {Sato}, \citenamefont {Okamoto}, \citenamefont {Tserkovnyak},
  \citenamefont {Qiu},\ and\ \citenamefont {Saitoh}}]{qiu}%
  \BibitemOpen
  \bibfield  {author} {\bibinfo {author} {\bibfnamefont {Z.}~\bibnamefont
  {Qiu}}, \bibinfo {author} {\bibfnamefont {J.}~\bibnamefont {Li}}, \bibinfo
  {author} {\bibfnamefont {D.}~\bibnamefont {Hou}}, \bibinfo {author}
  {\bibfnamefont {E.}~\bibnamefont {Arenholz}}, \bibinfo {author}
  {\bibfnamefont {A.~T.}\ \bibnamefont {N'Diaye}}, \bibinfo {author}
  {\bibfnamefont {A.}~\bibnamefont {Tan}}, \bibinfo {author} {\bibfnamefont
  {K.}~\bibnamefont {ichi Uchida}}, \bibinfo {author} {\bibfnamefont
  {K.}~\bibnamefont {Sato}}, \bibinfo {author} {\bibfnamefont {S.}~\bibnamefont
  {Okamoto}}, \bibinfo {author} {\bibfnamefont {Y.}~\bibnamefont
  {Tserkovnyak}}, \bibinfo {author} {\bibfnamefont {Z.~Q.}\ \bibnamefont
  {Qiu}}, \ and\ \bibinfo {author} {\bibfnamefont {E.}~\bibnamefont {Saitoh}},\
  }\href@noop {} {\bibfield  {journal} {\bibinfo  {journal} {Nat. Commun.}\
  }\textbf {\bibinfo {volume} {7}},\ \bibinfo {pages} {12670} (\bibinfo {year}
  {2016})}\BibitemShut {NoStop}%
\bibitem [{\citenamefont {Han}\ \emph {et~al.}(2014)\citenamefont {Han},
  \citenamefont {Song}, \citenamefont {F.~Li}, \citenamefont {Wang},
  \citenamefont {Yang},\ and\ \citenamefont {Pan}}]{jhhan}%
  \BibitemOpen
  \bibfield  {author} {\bibinfo {author} {\bibfnamefont {J.~H.}\ \bibnamefont
  {Han}}, \bibinfo {author} {\bibfnamefont {C.}~\bibnamefont {Song}}, \bibinfo
  {author} {\bibfnamefont {Y.~Y.~W.}\ \bibnamefont {F.~Li}}, \bibinfo {author}
  {\bibfnamefont {G.~Y.}\ \bibnamefont {Wang}}, \bibinfo {author}
  {\bibfnamefont {Q.~H.}\ \bibnamefont {Yang}}, \ and\ \bibinfo {author}
  {\bibfnamefont {F.}~\bibnamefont {Pan}},\ }\href@noop {} {\bibfield
  {journal} {\bibinfo  {journal} {Phys. Rev. B}\ }\textbf {\bibinfo {volume}
  {90}},\ \bibinfo {pages} {144431} (\bibinfo {year} {2014})}\BibitemShut
  {NoStop}%
\bibitem [{\citenamefont {Hoogeboom}\ \emph {et~al.}()\citenamefont
  {Hoogeboom}, \citenamefont {Aqeel}, \citenamefont {Kuschel}, \citenamefont
  {Palstra},\ and\ \citenamefont {van Wees}}]{hoogeboom}%
  \BibitemOpen
  \bibfield  {author} {\bibinfo {author} {\bibfnamefont {G.~R.}\ \bibnamefont
  {Hoogeboom}}, \bibinfo {author} {\bibfnamefont {A.}~\bibnamefont {Aqeel}},
  \bibinfo {author} {\bibfnamefont {T.}~\bibnamefont {Kuschel}}, \bibinfo
  {author} {\bibfnamefont {T.~T.~M.}\ \bibnamefont {Palstra}}, \ and\ \bibinfo
  {author} {\bibfnamefont {B.~J.}\ \bibnamefont {van Wees}},\ }\href@noop {}
  {\bibinfo  {journal} {arXiv:1706.03004}\ }\BibitemShut {NoStop}%
\bibitem [{\citenamefont {H\"{o}lzle}(1993)}]{Holzle}%
  \BibitemOpen
\bibfield  {journal} {  }\bibfield  {author} {\bibinfo {author} {\bibfnamefont
  {R.}~\bibnamefont {H\"{o}lzle}},\ }\href@noop {} {\emph {\bibinfo {title}
  {Magnetismus von Festk\"{o}rpern und Grenzfl\"{a}chen}}}\ (\bibinfo
  {publisher} {Forschungszentrum Jülich},\ \bibinfo {year} {1993})\BibitemShut
  {NoStop}%
\bibitem [{\citenamefont {Fina}\ \emph {et~al.}(2014)\citenamefont {Fina},
  \citenamefont {Marti}, \citenamefont {Yi}, \citenamefont {Liu}, \citenamefont
  {Chu}, \citenamefont {Rayan-Serrao}, \citenamefont {Suresha}, \citenamefont
  {Shick}, \citenamefont {\v{Z}elezn\'{y}}, \citenamefont {Jungwirth},
  \citenamefont {Fontcuberta},\ and\ \citenamefont {Ramesh}}]{fina}%
  \BibitemOpen
  \bibfield  {author} {\bibinfo {author} {\bibfnamefont {I.}~\bibnamefont
  {Fina}}, \bibinfo {author} {\bibfnamefont {X.}~\bibnamefont {Marti}},
  \bibinfo {author} {\bibfnamefont {D.}~\bibnamefont {Yi}}, \bibinfo {author}
  {\bibfnamefont {J.}~\bibnamefont {Liu}}, \bibinfo {author} {\bibfnamefont
  {J.}~\bibnamefont {Chu}}, \bibinfo {author} {\bibfnamefont {C.}~\bibnamefont
  {Rayan-Serrao}}, \bibinfo {author} {\bibfnamefont {S.}~\bibnamefont
  {Suresha}}, \bibinfo {author} {\bibfnamefont {A.}~\bibnamefont {Shick}},
  \bibinfo {author} {\bibfnamefont {J.}~\bibnamefont {\v{Z}elezn\'{y}}},
  \bibinfo {author} {\bibfnamefont {T.}~\bibnamefont {Jungwirth}}, \bibinfo
  {author} {\bibfnamefont {J.}~\bibnamefont {Fontcuberta}}, \ and\ \bibinfo
  {author} {\bibfnamefont {R.}~\bibnamefont {Ramesh}},\ }\href@noop {}
  {\bibfield  {journal} {\bibinfo  {journal} {Nat. Commun.}\ }\textbf {\bibinfo
  {volume} {5}},\ \bibinfo {pages} {4671} (\bibinfo {year} {2014})}\BibitemShut
  {NoStop}%
\bibitem [{\citenamefont {Wu}\ \emph {et~al.}(2016{\natexlab{b}})\citenamefont
  {Wu}, \citenamefont {Abid}, \citenamefont {Kalitsov}, \citenamefont
  {Zarzhitsky}, \citenamefont {Abid}, \citenamefont {Liao}, \citenamefont
  {Coile\'{a}in}, \citenamefont {Xu}, \citenamefont {Wang}, \citenamefont
  {Liu}, \citenamefont {Mryasov}, \citenamefont {Chang},\ and\ \citenamefont
  {Shvets}}]{hanchun}%
  \BibitemOpen
  \bibfield  {author} {\bibinfo {author} {\bibfnamefont {H.-C.}\ \bibnamefont
  {Wu}}, \bibinfo {author} {\bibfnamefont {M.}~\bibnamefont {Abid}}, \bibinfo
  {author} {\bibfnamefont {A.}~\bibnamefont {Kalitsov}}, \bibinfo {author}
  {\bibfnamefont {P.}~\bibnamefont {Zarzhitsky}}, \bibinfo {author}
  {\bibfnamefont {M.}~\bibnamefont {Abid}}, \bibinfo {author} {\bibfnamefont
  {Z.-M.}\ \bibnamefont {Liao}}, \bibinfo {author} {\bibfnamefont {C.~O.}\
  \bibnamefont {Coile\'{a}in}}, \bibinfo {author} {\bibfnamefont
  {H.}~\bibnamefont {Xu}}, \bibinfo {author} {\bibfnamefont {J.-J.}\
  \bibnamefont {Wang}}, \bibinfo {author} {\bibfnamefont {H.}~\bibnamefont
  {Liu}}, \bibinfo {author} {\bibfnamefont {O.~N.}\ \bibnamefont {Mryasov}},
  \bibinfo {author} {\bibfnamefont {C.-R.}\ \bibnamefont {Chang}}, \ and\
  \bibinfo {author} {\bibfnamefont {I.~V.}\ \bibnamefont {Shvets}},\
  }\href@noop {} {\bibfield  {journal} {\bibinfo  {journal} {Advanced
  Functional Materials}\ }\textbf {\bibinfo {volume} {26}},\ \bibinfo {pages}
  {5884} (\bibinfo {year} {2016}{\natexlab{b}})}\BibitemShut {NoStop}%
\bibitem [{\citenamefont {G\"{a}fvert}\ \emph {et~al.}(1977)\citenamefont
  {G\"{a}fvert}, \citenamefont {Lundgren}, \citenamefont {Westerstrandh},\ and\
  \citenamefont {Beckman}}]{Gafvert}%
  \BibitemOpen
  \bibfield  {author} {\bibinfo {author} {\bibfnamefont {U.}~\bibnamefont
  {G\"{a}fvert}}, \bibinfo {author} {\bibfnamefont {L.}~\bibnamefont
  {Lundgren}}, \bibinfo {author} {\bibfnamefont {B.}~\bibnamefont
  {Westerstrandh}}, \ and\ \bibinfo {author} {\bibfnamefont {O.}~\bibnamefont
  {Beckman}},\ }\href@noop {} {\bibfield  {journal} {\bibinfo  {journal} {J.
  Phys. Chem. Solids}\ }\textbf {\bibinfo {volume} {38}},\ \bibinfo {pages}
  {1333} (\bibinfo {year} {1977})}\BibitemShut {NoStop}%
\bibitem [{\citenamefont {Foner}(1963)}]{Foner}%
  \BibitemOpen
  \bibfield  {author} {\bibinfo {author} {\bibfnamefont {S.}~\bibnamefont
  {Foner}},\ }\href@noop {} {\bibfield  {journal} {\bibinfo  {journal} {Phys.
  Rev.}\ }\textbf {\bibinfo {volume} {130}},\ \bibinfo {pages} {183} (\bibinfo
  {year} {1963})}\BibitemShut {NoStop}%
\bibitem [{\citenamefont {Beckman}\ \emph {et~al.}(1968)\citenamefont
  {Beckman}, \citenamefont {Bruckner}, \citenamefont {Fuchs}, \citenamefont
  {Ritter},\ and\ \citenamefont {Wegener}}]{Beckman}%
  \BibitemOpen
  \bibfield  {author} {\bibinfo {author} {\bibfnamefont {V.}~\bibnamefont
  {Beckman}}, \bibinfo {author} {\bibfnamefont {W.}~\bibnamefont {Bruckner}},
  \bibinfo {author} {\bibfnamefont {W.}~\bibnamefont {Fuchs}}, \bibinfo
  {author} {\bibfnamefont {G.}~\bibnamefont {Ritter}}, \ and\ \bibinfo {author}
  {\bibfnamefont {H.}~\bibnamefont {Wegener}},\ }\href@noop {} {\bibfield
  {journal} {\bibinfo  {journal} {phys. stat. sol.}\ }\textbf {\bibinfo
  {volume} {29}},\ \bibinfo {pages} {781} (\bibinfo {year} {1968})}\BibitemShut
  {NoStop}%
\bibitem [{\citenamefont {Rebbouh}\ \emph {et~al.}(2007)\citenamefont
  {Rebbouh}, \citenamefont {Hermann},\ and\ \citenamefont
  {Grandjean}}]{Rebbouh}%
  \BibitemOpen
  \bibfield  {author} {\bibinfo {author} {\bibfnamefont {L.}~\bibnamefont
  {Rebbouh}}, \bibinfo {author} {\bibfnamefont {R.~P.}\ \bibnamefont
  {Hermann}}, \ and\ \bibinfo {author} {\bibfnamefont {F.}~\bibnamefont
  {Grandjean}},\ }\href@noop {} {\bibfield  {journal} {\bibinfo  {journal}
  {Phys. Rev. B.}\ }\textbf {\bibinfo {volume} {76}},\ \bibinfo {pages}
  {174422} (\bibinfo {year} {2007})}\BibitemShut {NoStop}%
\bibitem [{\citenamefont {Nakayama}\ \emph {et~al.}(2013)\citenamefont
  {Nakayama}, \citenamefont {Althammer}, \citenamefont {Chen}, \citenamefont
  {Uchida}, \citenamefont {Kajiwara}, \citenamefont {Kikuchi}, \citenamefont
  {Ohtani}, \citenamefont {Gepr\"{a}gs}, \citenamefont {Opel}, \citenamefont
  {Takahashi}, \citenamefont {Gross}, \citenamefont {Bauer}, \citenamefont
  {Goennenwein},\ and\ \citenamefont {Saitoh}}]{Nakayama}%
  \BibitemOpen
  \bibfield  {author} {\bibinfo {author} {\bibfnamefont {H.}~\bibnamefont
  {Nakayama}}, \bibinfo {author} {\bibfnamefont {M.}~\bibnamefont {Althammer}},
  \bibinfo {author} {\bibfnamefont {Y.~T.}\ \bibnamefont {Chen}}, \bibinfo
  {author} {\bibfnamefont {K.}~\bibnamefont {Uchida}}, \bibinfo {author}
  {\bibfnamefont {Y.}~\bibnamefont {Kajiwara}}, \bibinfo {author}
  {\bibfnamefont {D.}~\bibnamefont {Kikuchi}}, \bibinfo {author} {\bibfnamefont
  {T.}~\bibnamefont {Ohtani}}, \bibinfo {author} {\bibfnamefont
  {S.}~\bibnamefont {Gepr\"{a}gs}}, \bibinfo {author} {\bibfnamefont
  {M.}~\bibnamefont {Opel}}, \bibinfo {author} {\bibfnamefont {S.}~\bibnamefont
  {Takahashi}}, \bibinfo {author} {\bibfnamefont {R.}~\bibnamefont {Gross}},
  \bibinfo {author} {\bibfnamefont {G.~E.~W.}\ \bibnamefont {Bauer}}, \bibinfo
  {author} {\bibfnamefont {S.~T.~B.}\ \bibnamefont {Goennenwein}}, \ and\
  \bibinfo {author} {\bibfnamefont {E.}~\bibnamefont {Saitoh}},\ }\href@noop {}
  {\bibfield  {journal} {\bibinfo  {journal} {Phys. Rev. Lett.}\ }\textbf
  {\bibinfo {volume} {110}},\ \bibinfo {pages} {206601} (\bibinfo {year}
  {2013})}\BibitemShut {NoStop}%
\bibitem [{\citenamefont {Brataas}\ \emph {et~al.}(2006)\citenamefont
  {Brataas}, \citenamefont {Bauer},\ and\ \citenamefont {Kelly}}]{Brataas}%
  \BibitemOpen
  \bibfield  {author} {\bibinfo {author} {\bibfnamefont {A.}~\bibnamefont
  {Brataas}}, \bibinfo {author} {\bibfnamefont {G.~E.~W.}\ \bibnamefont
  {Bauer}}, \ and\ \bibinfo {author} {\bibfnamefont {P.~J.}\ \bibnamefont
  {Kelly}},\ }\href@noop {} {\bibfield  {journal} {\bibinfo  {journal} {Phys.
  Rep.}\ }\textbf {\bibinfo {volume} {427}},\ \bibinfo {pages} {157} (\bibinfo
  {year} {2006})}\BibitemShut {NoStop}%
\bibitem [{\citenamefont {Takahashi}\ and\ \citenamefont
  {Maekawa}(2008)}]{takahashi}%
  \BibitemOpen
  \bibfield  {author} {\bibinfo {author} {\bibfnamefont {S.}~\bibnamefont
  {Takahashi}}\ and\ \bibinfo {author} {\bibfnamefont {S.}~\bibnamefont
  {Maekawa}},\ }\href@noop {} {\bibfield  {journal} {\bibinfo  {journal} {J.
  Phys. Soc. Jpn.}\ }\textbf {\bibinfo {volume} {77}},\ \bibinfo {pages}
  {031009} (\bibinfo {year} {2008})}\BibitemShut {NoStop}%
\bibitem [{\citenamefont {Chen}\ \emph {et~al.}(2013)\citenamefont {Chen},
  \citenamefont {Takahashi}, \citenamefont {Nakayama}, \citenamefont
  {M.Althammer}, \citenamefont {Goennenwein}, \citenamefont {Saitoh},\ and\
  \citenamefont {Bauer}}]{yan}%
  \BibitemOpen
  \bibfield  {author} {\bibinfo {author} {\bibfnamefont {Y.~T.}\ \bibnamefont
  {Chen}}, \bibinfo {author} {\bibfnamefont {S.}~\bibnamefont {Takahashi}},
  \bibinfo {author} {\bibfnamefont {H.}~\bibnamefont {Nakayama}}, \bibinfo
  {author} {\bibnamefont {M.Althammer}}, \bibinfo {author} {\bibfnamefont
  {S.~T.~B.}\ \bibnamefont {Goennenwein}}, \bibinfo {author} {\bibfnamefont
  {E.}~\bibnamefont {Saitoh}}, \ and\ \bibinfo {author} {\bibfnamefont
  {G.~E.~W.}\ \bibnamefont {Bauer}},\ }\href@noop {} {\bibfield  {journal}
  {\bibinfo  {journal} {Phys. Rev. B}\ }\textbf {\bibinfo {volume} {87}},\
  \bibinfo {pages} {144411} (\bibinfo {year} {2013})}\BibitemShut {NoStop}%
\bibitem [{\citenamefont {Kondoh}\ \emph {et~al.}(1958)\citenamefont {Kondoh},
  \citenamefont {Uchida}, \citenamefont {Nakazumi},\ and\ \citenamefont
  {Nagamiya}}]{kondoh}%
  \BibitemOpen
  \bibfield  {author} {\bibinfo {author} {\bibfnamefont {H.}~\bibnamefont
  {Kondoh}}, \bibinfo {author} {\bibfnamefont {E.}~\bibnamefont {Uchida}},
  \bibinfo {author} {\bibfnamefont {Y.}~\bibnamefont {Nakazumi}}, \ and\
  \bibinfo {author} {\bibfnamefont {T.}~\bibnamefont {Nagamiya}},\ }\href@noop
  {} {\bibfield  {journal} {\bibinfo  {journal} {J. Phys. Soc. Jpn.}\ }\textbf
  {\bibinfo {volume} {13}},\ \bibinfo {pages} {579} (\bibinfo {year}
  {1958})}\BibitemShut {NoStop}%
\bibitem [{\citenamefont {Uchida}\ \emph {et~al.}(1964)\citenamefont {Uchida},
  \citenamefont {Fukuoka}, \citenamefont {Kondoh}, \citenamefont {Takeda},
  \citenamefont {Nakazumi},\ and\ \citenamefont {Nagamiya}}]{uchida1}%
  \BibitemOpen
  \bibfield  {author} {\bibinfo {author} {\bibfnamefont {E.}~\bibnamefont
  {Uchida}}, \bibinfo {author} {\bibfnamefont {N.}~\bibnamefont {Fukuoka}},
  \bibinfo {author} {\bibfnamefont {H.}~\bibnamefont {Kondoh}}, \bibinfo
  {author} {\bibfnamefont {T.}~\bibnamefont {Takeda}}, \bibinfo {author}
  {\bibfnamefont {Y.}~\bibnamefont {Nakazumi}}, \ and\ \bibinfo {author}
  {\bibfnamefont {T.}~\bibnamefont {Nagamiya}},\ }\href@noop {} {\bibfield
  {journal} {\bibinfo  {journal} {J. Phys. Soc. Jpn.}\ }\textbf {\bibinfo
  {volume} {19}},\ \bibinfo {pages} {2088} (\bibinfo {year}
  {1964})}\BibitemShut {NoStop}%
\bibitem [{\citenamefont {Hirsch}(1999)}]{hirsch}%
  \BibitemOpen
  \bibfield  {author} {\bibinfo {author} {\bibfnamefont {J.~E.}\ \bibnamefont
  {Hirsch}},\ }\href@noop {} {\bibfield  {journal} {\bibinfo  {journal} {Phys.
  Rev. Lett.}\ }\textbf {\bibinfo {volume} {83}},\ \bibinfo {pages} {1834}
  (\bibinfo {year} {1999})}\BibitemShut {NoStop}%
\bibitem [{\citenamefont {Valenzuela}\ and\ \citenamefont
  {Tinkham}(2006)}]{valen}%
  \BibitemOpen
  \bibfield  {author} {\bibinfo {author} {\bibfnamefont {S.~O.}\ \bibnamefont
  {Valenzuela}}\ and\ \bibinfo {author} {\bibfnamefont {M.}~\bibnamefont
  {Tinkham}},\ }\href@noop {} {\bibfield  {journal} {\bibinfo  {journal}
  {Nature}\ }\textbf {\bibinfo {volume} {442}},\ \bibinfo {pages} {176}
  (\bibinfo {year} {2006})}\BibitemShut {NoStop}%
\bibitem [{\citenamefont {Sinova}\ \emph {et~al.}(2015)\citenamefont {Sinova},
  \citenamefont {Valenzuela}, \citenamefont {Wunderlich}, \citenamefont
  {Back},\ and\ \citenamefont {Jungwirth}}]{sinova}%
  \BibitemOpen
  \bibfield  {author} {\bibinfo {author} {\bibfnamefont {J.}~\bibnamefont
  {Sinova}}, \bibinfo {author} {\bibfnamefont {S.~O.}\ \bibnamefont
  {Valenzuela}}, \bibinfo {author} {\bibfnamefont {J.}~\bibnamefont
  {Wunderlich}}, \bibinfo {author} {\bibfnamefont {C.~H.}\ \bibnamefont
  {Back}}, \ and\ \bibinfo {author} {\bibfnamefont {T.}~\bibnamefont
  {Jungwirth}},\ }\href@noop {} {\bibfield  {journal} {\bibinfo  {journal}
  {Rev. Mod. Phys.}\ }\textbf {\bibinfo {volume} {87}},\ \bibinfo {pages}
  {1213} (\bibinfo {year} {2015})}\BibitemShut {NoStop}%
\bibitem [{\citenamefont {Saitoh}\ \emph {et~al.}(2006)\citenamefont {Saitoh},
  \citenamefont {Ueda},\ and\ \citenamefont {Miyajima}}]{saitoh}%
  \BibitemOpen
  \bibfield  {author} {\bibinfo {author} {\bibfnamefont {E.}~\bibnamefont
  {Saitoh}}, \bibinfo {author} {\bibfnamefont {M.}~\bibnamefont {Ueda}}, \ and\
  \bibinfo {author} {\bibfnamefont {H.}~\bibnamefont {Miyajima}},\ }\href@noop
  {} {\bibfield  {journal} {\bibinfo  {journal} {Appl. Phys. Lett.}\ }\textbf
  {\bibinfo {volume} {88}},\ \bibinfo {pages} {182509} (\bibinfo {year}
  {2006})}\BibitemShut {NoStop}%
\bibitem [{\citenamefont {Kittel}(1951)}]{Kittel}%
  \BibitemOpen
  \bibfield  {author} {\bibinfo {author} {\bibfnamefont {C.}~\bibnamefont
  {Kittel}},\ }\href@noop {} {\bibfield  {journal} {\bibinfo  {journal} {Phys.
  Rev.}\ }\textbf {\bibinfo {volume} {82}},\ \bibinfo {pages} {565} (\bibinfo
  {year} {1951})}\BibitemShut {NoStop}%
\bibitem [{\citenamefont {Keffer}\ and\ \citenamefont {Kittel}(1952)}]{Keffer}%
  \BibitemOpen
  \bibfield  {author} {\bibinfo {author} {\bibfnamefont {F.}~\bibnamefont
  {Keffer}}\ and\ \bibinfo {author} {\bibfnamefont {C.}~\bibnamefont
  {Kittel}},\ }\href@noop {} {\bibfield  {journal} {\bibinfo  {journal} {Phys.
  Rev.}\ }\textbf {\bibinfo {volume} {85}},\ \bibinfo {pages} {329} (\bibinfo
  {year} {1952})}\BibitemShut {NoStop}%
\bibitem [{\citenamefont {Nagamiya}(1951)}]{Nagamiya}%
  \BibitemOpen
  \bibfield  {author} {\bibinfo {author} {\bibfnamefont {T.}~\bibnamefont
  {Nagamiya}},\ }\href@noop {} {\bibfield  {journal} {\bibinfo  {journal}
  {Progr. Theoret. Phys. (Kyoto)}\ }\textbf {\bibinfo {volume} {6}},\ \bibinfo
  {pages} {350} (\bibinfo {year} {1951})}\BibitemShut {NoStop}%
\bibitem [{\citenamefont {Nagamiya}\ \emph {et~al.}(1955)\citenamefont
  {Nagamiya}, \citenamefont {Yosida},\ and\ \citenamefont {Kubo}}]{Nagamiya2}%
  \BibitemOpen
  \bibfield  {author} {\bibinfo {author} {\bibfnamefont {T.}~\bibnamefont
  {Nagamiya}}, \bibinfo {author} {\bibfnamefont {K.}~\bibnamefont {Yosida}}, \
  and\ \bibinfo {author} {\bibfnamefont {R.}~\bibnamefont {Kubo}},\ }\href@noop
  {} {\bibfield  {journal} {\bibinfo  {journal} {Advances in Physics}\ }\textbf
  {\bibinfo {volume} {4}},\ \bibinfo {pages} {1} (\bibinfo {year}
  {1955})}\BibitemShut {NoStop}%
\bibitem [{\citenamefont {Bogdanov}\ \emph {et~al.}(2007)\citenamefont
  {Bogdanov}, \citenamefont {Zhuravlev},\ and\ \citenamefont
  {R\"{o}{\ss}ler}}]{Bogdanov}%
  \BibitemOpen
  \bibfield  {author} {\bibinfo {author} {\bibfnamefont {A.~N.}\ \bibnamefont
  {Bogdanov}}, \bibinfo {author} {\bibfnamefont {A.~V.}\ \bibnamefont
  {Zhuravlev}}, \ and\ \bibinfo {author} {\bibfnamefont {U.~K.}\ \bibnamefont
  {R\"{o}{\ss}ler}},\ }\href@noop {} {\bibfield  {journal} {\bibinfo  {journal}
  {Phys. Rev. B}\ }\textbf {\bibinfo {volume} {75}},\ \bibinfo {pages} {094425}
  (\bibinfo {year} {2007})}\BibitemShut {NoStop}%
\bibitem [{\citenamefont {Filippetti}\ and\ \citenamefont
  {Fiorentini}(2005)}]{ales}%
  \BibitemOpen
  \bibfield  {author} {\bibinfo {author} {\bibfnamefont {A.}~\bibnamefont
  {Filippetti}}\ and\ \bibinfo {author} {\bibfnamefont {V.}~\bibnamefont
  {Fiorentini}},\ }\href@noop {} {\bibfield  {journal} {\bibinfo  {journal}
  {Phys. Rev. Lett.}\ }\textbf {\bibinfo {volume} {95}},\ \bibinfo {pages}
  {086405} (\bibinfo {year} {2005})}\BibitemShut {NoStop}%
\bibitem [{\citenamefont {van Schilfgaarde}\ and\ \citenamefont
  {Antropov}(1999)}]{van}%
  \BibitemOpen
  \bibfield  {author} {\bibinfo {author} {\bibfnamefont {M.}~\bibnamefont {van
  Schilfgaarde}}\ and\ \bibinfo {author} {\bibfnamefont {V.~P.}\ \bibnamefont
  {Antropov}},\ }\href@noop {} {\bibfield  {journal} {\bibinfo  {journal}
  {Journal of Applied Physics}\ }\textbf {\bibinfo {volume} {85}},\ \bibinfo
  {pages} {4827} (\bibinfo {year} {1999})}\BibitemShut {NoStop}%
\bibitem [{\citenamefont {Pchelkina}\ and\ \citenamefont
  {Solovyev}(2015)}]{pch}%
  \BibitemOpen
  \bibfield  {author} {\bibinfo {author} {\bibfnamefont {Z.~V.}\ \bibnamefont
  {Pchelkina}}\ and\ \bibinfo {author} {\bibfnamefont {I.~V.}\ \bibnamefont
  {Solovyev}},\ }\href@noop {} {\bibfield  {journal} {\bibinfo  {journal}
  {Journal of Physics: Condensed Matter}\ }\textbf {\bibinfo {volume} {27}},\
  \bibinfo {pages} {026001} (\bibinfo {year} {2015})}\BibitemShut {NoStop}%
\bibitem [{\citenamefont {Uchida}\ \emph {et~al.}(1967)\citenamefont {Uchida},
  \citenamefont {Fukuoka}, \citenamefont {Kondoh}, \citenamefont {Takeda},
  \citenamefont {Nakazumi},\ and\ \citenamefont {Nagamiya}}]{uchida2}%
  \BibitemOpen
  \bibfield  {author} {\bibinfo {author} {\bibfnamefont {E.}~\bibnamefont
  {Uchida}}, \bibinfo {author} {\bibfnamefont {N.}~\bibnamefont {Fukuoka}},
  \bibinfo {author} {\bibfnamefont {H.}~\bibnamefont {Kondoh}}, \bibinfo
  {author} {\bibfnamefont {T.}~\bibnamefont {Takeda}}, \bibinfo {author}
  {\bibfnamefont {Y.}~\bibnamefont {Nakazumi}}, \ and\ \bibinfo {author}
  {\bibfnamefont {T.}~\bibnamefont {Nagamiya}},\ }\href@noop {} {\bibfield
  {journal} {\bibinfo  {journal} {J. Phys. Soc. Jpn.}\ }\textbf {\bibinfo
  {volume} {23}},\ \bibinfo {pages} {1197} (\bibinfo {year}
  {1967})}\BibitemShut {NoStop}%
\bibitem [{\citenamefont {Inagawa}\ \emph {et~al.}(1971)\citenamefont
  {Inagawa}, \citenamefont {Kamigaki},\ and\ \citenamefont {Miura}}]{inagawa}%
  \BibitemOpen
  \bibfield  {author} {\bibinfo {author} {\bibfnamefont {K.}~\bibnamefont
  {Inagawa}}, \bibinfo {author} {\bibfnamefont {K.}~\bibnamefont {Kamigaki}}, \
  and\ \bibinfo {author} {\bibfnamefont {S.}~\bibnamefont {Miura}},\
  }\href@noop {} {\bibfield  {journal} {\bibinfo  {journal} {J. Phys. Soc.
  Jpn.}\ }\textbf {\bibinfo {volume} {31}},\ \bibinfo {pages} {1276} (\bibinfo
  {year} {1971})}\BibitemShut {NoStop}%
\end{thebibliography}

\end{document}